\documentclass[11pt,a4paper]{article}
\usepackage{jheppub}
\usepackage{indentfirst}
\usepackage{amsmath}
\usepackage{bigstrut}
\usepackage{amssymb}
\usepackage{array}
\usepackage{multirow}
\usepackage{floatrow}
\usepackage{etoolbox}
\usepackage{graphicx}
\usepackage{hyperref}
\usepackage{amsmath}
\usepackage{amsthm}
\usepackage{slashed}
\usepackage{array}
\usepackage{epstopdf}
\usepackage{graphicx}
\usepackage{fancyhdr}
\usepackage{float}
\usepackage{caption}
\usepackage{subcaption}
\usepackage[dvipsnames]{xcolor}
\usepackage{hyperref}
\usepackage[utf8]{inputenc}
\usepackage[T1]{fontenc}
\usepackage{lmodern}
\usepackage{textcomp}
\usepackage{microtype}
\usepackage[compat=1.0.0]{tikz-feynman}
\usepackage{wasysym}
\usepackage{hyperref}
\usepackage{mathtools}
\usepackage{physics}
\usepackage{units}
\usepackage{siunitx}
\usepackage{ulem}
\usepackage{scrextend}
\graphicspath{{./}{Figs/}}

\newcommand{\eq}[1]{Eq.~(\ref{eq:#1})}
\newcommand{\fig}[1]{Fig.~(\ref{fig:#1})}
\newcommand{\tab}[1]{Tab.~(\ref{tab:#1})}
\newcommand{\sect}[1]{Sec.~\ref{sec:#1}}
\newcommand{\app}[1]{Appendix.~\ref{app:#1}}

\newcommand{\zprime}[0]{Z^{\prime}}

\preprint{FERMILAB-PUB-22-766-T}
\title{Anatomy of the Electroweak Phase Transition  for Dark Sector induced Baryogenesis }
 
\author[a,b,c]{Marcela Carena,}
\affiliation[a]{Theoretical Physics Department, Fermi National Accelerator Laboratory, Batavia, Illinois, 60510, USA}
\affiliation[b]{Enrico Fermi Institute, University of Chicago, Chicago, Illinois, 60637, USA}
\affiliation[c]{Kavli Institute for Cosmological Physics, University of Chicago, Chicago, Illinois, 60637,
USA}
\emailAdd{carena@fnal.gov}
\author[a]{Ying-Ying Li, }
\emailAdd{yingyingli1013@outlook.com}
\author[a,b]{Tong Ou, }
\emailAdd{tongou@uchicago.edu}
\author[d]{Yikun Wang}
\affiliation[d]{Walter Burke Institute for Theoretical Physics, California Institute of Technology, Pasadena, CA 91125}
\emailAdd{yikunw@caltech.edu}

\abstract{We investigate the electroweak phase transition patterns for a recently proposed baryogenesis model
with CP violation originated in the dark sector. The model   includes a complex scalar singlet-Higgs boson portal, a  $U(1)_l$ gauge lepton symmetry  with a $\zprime$ gauge boson portal and  a fermionic dark matter particle. We find a novel thermal history of the scalar sector, featuring a $Z_2$ breaking singlet vacuum in the early Universe driven by a dark Yukawa coupling, that induces  a one-step strongly first order electroweak phase transition. 
We explore the parameter space that generates the observed matter-antimatter asymmetry and dark matter relic abundance, while being consistent with constraints from electric dipole moment, collider searches, and dark matter direct detection bounds. The complex singlet can be produced via the Higgs portal and decays into Standard Model particles after traveling a certain distance. We explore the reach for long-lived singlet scalars at the  {\unit[13]{TeV}} Large Hadron Collider with $\mathcal{L}=\unit[139]{fb^{-1}}$ and show its impact on the parameter space of the model. 
Setting aside currently unresolved theoretical uncertainties, we estimate the gravitational wave signatures detectable at future observatories.
}

\date{\today}

\begin{document}
\maketitle
\unitlength = 1mm
\section{Introduction} 
Electroweak baryogenesis (EWBG) is an elegant mechanism that generates the observed baryon asymmetry in the Universe (BAU) at the electroweak phase transition (EWPT) \cite{KUZMIN198536,COHEN1990561}. The Standard Model (SM), however, can not account for successful EWBG: the Higgs mass is too heavy to allow for a strong first order phase transition and the CP violation source is not sufficient, hence new physics is required. 
However, new sources of CP violation from particles charged under the SM are strongly constrained by the remarkable results from electric dipole moment (EDM) experiments \cite{andreev2018improved,269}. Recently, a new EWBG mechanism in which CP violation occurs in the dark sector and is transmitted to the visible sector via a vector gauge boson $\zprime$ from a $U(1)_l$ gauge lepton symmetry has been proposed \cite{carena2019electroweak, carena2020dark}. In such a scenario, a complex scalar singlet $S$, provides the source of CP violation through its Yukawa coupling to a dark fermion $\chi$ charged under $U(1)_l$ and leads to a strongly first order electroweak phase transition (SFOEWPT) via its coupling with the Higgs boson field. In this way, the contribution to EDM is suppressed to beyond two-loop level and compatible with current experiments. The dark fermion $\chi$ can also serve as an ideal dark matter candidate. 

In this work, we study the pattern of the EWPT in the proposed new mechanism \cite{carena2019electroweak, carena2020dark}, and investigate the viable parameter space compatible with a successful EWBG, the observed dark matter relic abundance, and phenomenological bounds from dark matter direct detection and collider searches. We also explore potentially observable gravitational wave (GW) signatures.
Singlet extensions of the SM addressing the BAU generation and the EWPT have been extensively studied with focus on scalars heavier than half of the Higgs mass \cite{Cline:2012hg,Jiang:2015cwa,Curtin:2014jma,Huang:2016cjm,Chiang:2017nmu,Beniwal:2017eik}. There are also studies on EWPT for light scalars but leaving out the discussion of a complete EWBG model \cite{Kozaczuk:2019pet,Carena:2019une}. Here we compute, both analytically and numerically, the possibility of EWBG for a broad range of scalar masses,  from $\mathcal{O}(1)$ GeV to a few hundred GeV. More specifically, while a phase transition pattern for~EWBG was assumed in \cite{carena2019electroweak, carena2020dark}, we now perform a detailed study of the patterns of EWPT for the underlying model. This includes implementing the complete effective finite temperature potential at one loop order with appropriate thermal resummation and performing the nucleation calculation to assure the completeness of the phase transition.

In a broader context of the~SM singlet extension, the presence of a dark fermion sector has a particular impact on the scalar potential.
The Yukawa term between the singlet $S$ and a dark fermion 
breaks the $Z_2$ symmetry ($S \to -S$) explicitly at tree level, and, with a non-zero bare mass of the dark fermion,
contributes a tadpole term to the singlet potential at one loop level. To avoid the mixing between the $S$ and the SM Higgs, we introduce counterterms to impose the expectation value of $S$ at the electroweak vacuum at zero temperature to be zero up to one loop order. At finite temperature, thermal corrections from the dark Yukawa coupling will drive the vacuum of $S$ away from zero, leading to a distinct thermal history of the scalar sector: the Universe would go through a one-step first order phase transition from this $Z_2$ breaking singlet vacuum to the electroweak symmetry breaking vacuum at a lower temperature. The first order phase transition can readily be strong with a tree level barrier between the two vacua, yielding a successful EWBG and detectable GW signals. 

Furthermore, the additional singlet scalars have a rich phenomenology, that is to be updated and discussed in this work. 
If the~EW vacuum were to break the $Z_2$ symmetry along the singlet direction, 
the singlet scalars would mix with the SM Higgs and thus could be probed by Higgs boson and  electroweak precision measurements \cite{Carena:2018vpt}, and also by Higgs exotic decays when the singlet is light \cite{Kozaczuk:2019pet,Carena:2019une,Cepeda:2021rql}. In our study, the scalar potential has a $Z_2$ symmetry at zero temperature, and, in principle, the singlet would be stable \cite{Curtin:2014jma} and when below the Higgs decay threshold, it could be probed by Higgs invisible decay searches \cite{Kozaczuk:2019pet}. In our model, however, as the singlet carries $U(1)_l$ charge, the heavier singlet can decay via the $Z'$ portal to SM leptons promptly. The presence of the dark Yukawa coupling allows the lighter singlet to decay into SM particles through the dark matter $\chi$ loop and the $\zprime$ portal. Thus the decay width of the lighter singlet scalars is generically suppressed by the heavy fermion loop and the small $U(1)_l$ gauge coupling constrained by LEP bounds. This leads to distinct signatures for the lighter singlet to decay either in the tracker, muon chamber or outside the detector.  
We will investigate the reach for long-lived singlet
scalars at the $\unit[13]{TeV}$ LHC with $\mathcal{L}=\unit[139]{fb^{-1}}$ to probe the parameter space with successful EWBG.

This paper is organized as follows: in \sect{mod}, we introduce the scalar potential and review the basic elements of the model, including the specific mechanism of EWBG and its implications for dark matter phenomenology. In \sect{analytical_study}, we study the thermal history as well as the critical and nucleation behavior of the~EWPT with analytical calculations. We also perform  numerical scans for two benchmark scenarios, where viable parameter space for successful SFOEWPT  consistent with analytical evaluations is found. In \sect{BA}, we discuss the mechanism of baryogenesis and evaluate the parameter space for successful EWBG through numerical scan. In \sect{phe}, we re-evaluate the phenomenological discussions for dark matter direct detection, and present an opportunity for long-lived particle searches for the scalar singlets in the newly revealed parameter space. We also compute the GW signature of the model and show potential compatibility with~EWBG. We reserve \sect{conclusion} for our conclusions. We collect various technical aspects in the appendices.

\section{The EWBG model}

\label{sec:mod}

\begin{figure}[htbp]
\centering
\includegraphics[width=0.7\textwidth]{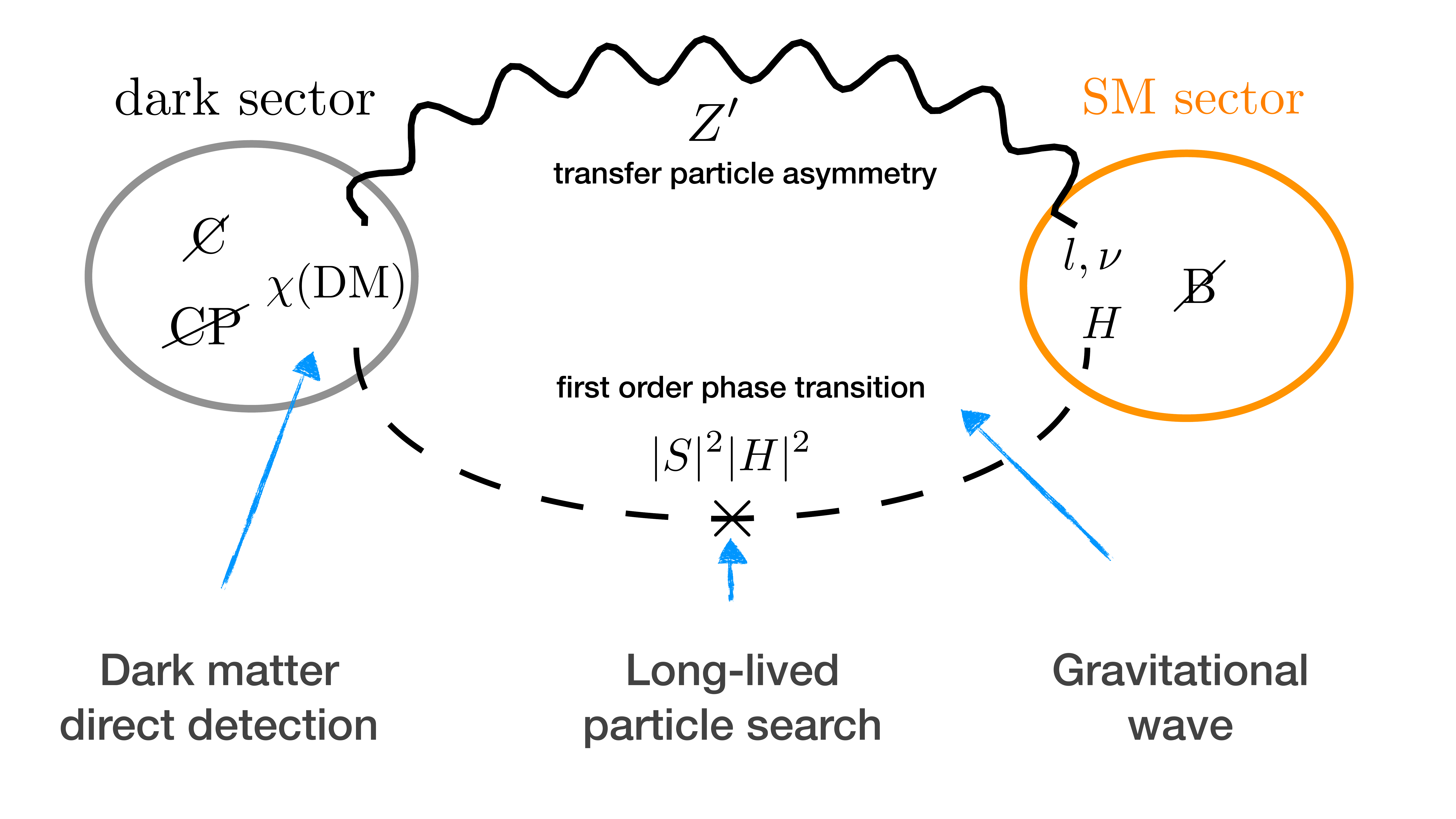}
\caption{Low energy sector of the model and probes for~EWBG and a dark matter candidate.}
\label{fig:model}
\end{figure}

In this section, we briefly introduce the model for EWBG as schematically shown in \fig{model}, with CP violation sourced from the dark sector, and an SFOEWPT where the barrier is provided by the tree-level coupling between the complex singlet and the SM~Higgs. In the following, we would present the Lagrangian  terms that define the Higgs and gauge boson portals between the SM sector and the dark sector of the theory. Readers can find more specific details of the model in \cite{carena2019electroweak, carena2020dark}.
In this work, we focus on the EW-scale low energy effective theory of the UV complete model for baryogenesis presented in  \cite{carena2019electroweak}, where the 
$U(1)_l$ symmetry, the gauge symmetry promoted from the lepton number, has been spontaneously broken by a heavy singlet field $\Phi$.
Integrating out various anomalon fields, the low energy sector of the model contains the~SM fields plus the new fields 
\begin{equation}
    Z_\mu',~~~S,~~~\chi_L, ~~~\chi_R.
\end{equation}
The complex singlet $S$ couples to the~SM Higgs to provide a barrier for the~EWPT. At zero temperature, the tree level effective scalar potential reads 
\begin{equation}
V_0 = \lambda_H(|H|^2-v_H^2)^2+\lambda_S(|S|^2-v_S^2)^2+\lambda_{SH}|S|^2|H|^2 + \delta V,
\label{eq:tree_t0}
\end{equation}
where $H$ is the Higgs doublet and 
\begin{equation} \label{eq:tree-dv}
 \delta V=\rho_S S+\kappa_S^2 S^2+\lambda_{3S}|S|^2 S+\text{h.c.}
\end{equation}
Observe that apart from the 
$\delta V$
 the tree level potential only depends on $|S|$.  Hence,  to accommodate a CP violating effect,  additional terms depending on $S$ are needed .
Such terms can arise from renormalizable, $U(1)_l$ invariant terms, involving the heavy singlet field $\Phi$. 
For this work, we study the case where $S$
has a zero vacuum expectation value (vev) at the EW vacuum, that implies no tadpole at zero temperature. This can be achieved by adjusting the coefficient for the tadpole term to be zero at 
 one-loop order,  as will be  discussed later. For simplicity, we will turn off the
cubic term. The coefficient of the quadratic term,  $\kappa_S^2$, is a  free parameter for scanning and is in general complex.

The complex SM $SU(2)$ Higgs doublet $H$ and the complex SM scalar singlet $S$ can be expressed in terms of real fields as follows
\begin{eqnarray}
\begin{aligned}
H&=\frac{1}{\sqrt{2}}\begin{pmatrix}
G_1+iG_2\\
h+iG_3
\end{pmatrix},~~
S&=\frac{1}{\sqrt{2}}(s+ia).
\end{aligned}
\end{eqnarray}
Using unitary gauge to eliminate the Goldstone fields $G_i$'s, we can rewrite the potential in \eq{tree_t0} in terms of $(h,s,a)$ as 
\begin{eqnarray}
V_0=&&-\frac{1}{2}\mu_H^2 h^2+\frac{1}{4}\lambda_H h^4 +\frac{1}{4}\lambda_{SH}h^2(s^2+a^2) \notag\\
&&-\frac{1}{2}\mu_S^2(s^2+a^2)+\frac{1}{4}\lambda_S(s^2+a^2)^2+\kappa_S^2(s^2-a^2),
\label{eq:tree_t0_cpvio}
\end{eqnarray}
where $\mu_{H(S)}^2=\lambda_{H(S)}v_{H(S)}^2$.
The Higgs quartic coupling and the Higgs vev at $T=0$ are fixed by Higgs boson measurements to be $\lambda_H=0.129$ and $v_H=246$ GeV. For convenience, we
rewrite the singlet quartic coupling $\lambda_{S}$ and the coefficient of the CP violating term $\kappa_S^2$ in
\eq{tree_t0_cpvio} in terms of physical parameters, i.e., the masses of the scalars $a,s$ at $T=0$, $m_{a,s}$:
\begin{equation}
m_a^2 = \frac{1}{2} \lambda_{SH} v_H^2 - \lambda_{S} v_S^{\prime 2},~~m_s^2 = m_a^2 + 4\kappa_S^2
\label{eq:scalarmass},
\end{equation}
with $v'^2_S=v_S^2+2\kappa_S^2/\lambda_S$. As will be shown in \sect{zero_temperature} and \sect{lightsingletTc}, $v_S^\prime$ is the value of the singlet field in the $a$-direction at a zero temperature EW symmetric stationary point of the scalar potential. The existence of this stationary point facilities a SFOEWPT.

The fermion fields $\chi_L$ and $\chi_R$ are SM singlets
that couple to the singlet composite scalar $S$. The Yukawa coupling between $S$ and $\chi$ is given by
\begin{equation}
\bar{\chi}_L(m_0+\lambda_c S)\chi_R+\text{h.c.}\, ,
\end{equation}
and the mass of $\chi$ reads
\begin{equation}
M_\chi=m_0+\lambda \exp(i[\theta+\text{arg}(S)])|S|,
\end{equation}
where $\lambda=|\lambda_c|$ and $\theta=\text{arg}(\lambda_c)$\footnote{Observe that in \cite{carena2019electroweak}, $\theta$ is defined as $\arg(\lambda_{c})+\arg(S)$ with $\arg(S)$ fixed to $\pi/2$.}. Note that $m_0$ is induced by the coupling of $\chi$ to the heavy singlet scalar field $\Phi$, when it acquires a non-zero vev at the~UV. The corresponding Yukawa coupling is assumed to be small such that $m_0$ is much lighter than the vev $\langle \Phi \rangle$ and the dark fermions remain dynamical at low energies. We use the freedom of field redefinition to make $\kappa_S^2$ and $m_0$ real and positive, leaving $\lambda_c$ complex in general. During the~EWPT, the  phase of $M_\chi$ varying in the direction of the expanding bubble wall, can be derived from $m_0$, the phases of  $\lambda_{c}$ and the  $S$ vev.
This is the physical source of CP violation, that will then induce a chiral asymmetry in the $\chi$ particles.

As has been mentioned above, the lepton number is promoted to a $U(1)_l$ gauge symmetry with an associated $Z'$ gauge boson, and the dark fermion $\chi$ and the singlet $S$ are assigned certain lepton number charges. Possible anomaly-free UV completions can be found in \cite{FileviezPerez:2011pt, Duerr:2013dza, Carena:2019xrr, Restrepo:2022cpq, FileviezPerez:2014lnj, FileviezPerez:2019cyn}. The new interactions introduced at low energy are:
\begin{eqnarray}
\mathcal{L} \supset g' Z'_\mu \left[(N_g + q)\bar\chi_L \gamma^\mu\chi_L + q \bar\chi_R \gamma^\mu\chi_R + \bar{L}\gamma^\mu L + \bar{e}_R\gamma^\mu e_R \right].
\end{eqnarray}
We assume $U(1)_l$ with $l=e,\mu,\tau$ and thus
$N_g=3$, and the charge  $q\sim\mathcal{O}(1)$ throughout this study. Given that $\chi_L$ and $\chi_R$ carry different $U(1)_l$ charges,  the chiral asymmetry in the $\chi$ sector will give a net $U(1)_l$ charge density near the bubble wall, that generates a background for the $Z_0^{\prime}$ component. Because of the coupling of the SM leptons with the $Z'$, this background  further generates a chemical potential for the SM leptons and consequently a net chiral asymmetry for them. As will be discussed in detail later on, solving the corresponding Boltzmann equation, considering the EW sphaleron rate suppressed inside the bubble wall, one obtains a lepton asymmetry. Ultimately, the sphaleron process, which preserves $B-L$, will generate equal asymmetry in the lepton and baryon sectors to source the observed BAU.

\begin{figure}[htbp]
\centering
\begin{subfigure}{0.32\textwidth}
\includegraphics[width=\textwidth]{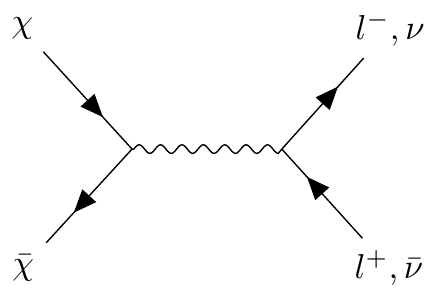}
\end{subfigure}
~~~~~~~~\begin{subfigure}{0.3\textwidth}
\includegraphics[width=0.7\textwidth]{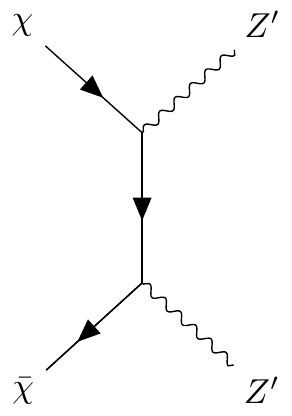}
\end{subfigure}
\begin{subfigure}{0.3\textwidth}
\includegraphics[width=0.7\textwidth]{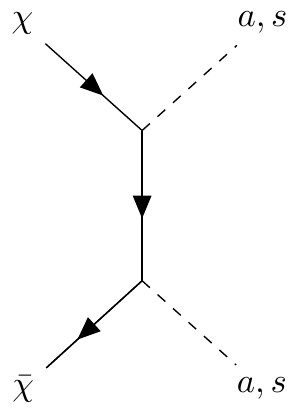}
\end{subfigure}
\caption{Tree-level annihilations of the dark fermion $\chi$.}
\label{fig:dmannihilation}
\end{figure}

Protected by a $Z_2$ symmetry in the Lagrangian, $\chi\to -\chi$, the dark fermion $\chi$ is stable and could be a dark matter candidate. The annihilation channels for $\chi$ at tree level include annihilating to $Z'Z'$, SM lepton pairs, $ss$, $aa$ and $sa$. The corresponding Feynman diagrams are shown in~\fig{dmannihilation}. Given the LEP constraints on $Z'$ search, the dominant annihilation channel to achieve the correct relic density is $\chi\bar{\chi}\rightarrow ss, aa, sa$, which requires \cite{carena2020dark}
\begin{equation}
\sqrt{\frac{m_0}{1.4\text{TeV}}}<\lambda<\sqrt{\frac{m_0}{1\text{TeV}}}.
\label{eq:dmtosinglets}
\end{equation}
The range for $\lambda$ is due to the $\theta$ dependence of the annihilation cross section. We will implement this relation in the numerical scanning to generate the observed dark matter relic abundance.

To conclude this section, we categorize our model parameters through the following groups:
\begin{eqnarray*}
\begin{aligned}
\rm fixed\ parameters&: \quad \lbrace \lambda_{H}, v_H \rbrace,\\
\rm scalar\ sector\ parameters&: \quad \{ \lambda_S, v'_S, \lambda_{SH}, m_a, m_s\},\\
\rm dark\ fermion\ parameters&: \quad \{ m_0,\lambda, \theta\},\\
\rm \textit{Z'}\ parameters&: \quad \{ g', M_{Z'}\}.
\end{aligned}
\end{eqnarray*}
The stability of the scalar potential and the EW thermal history would be affected by the fixed parameters and the scalar potential parameters at tree level, as well as  by the dark fermion parameters at loop level. The $Z'$ parameters do not enter the scalar potential but would be crucial for the transmission of the baryon asymmetry and thus are relevant for the~BAU. They are also of great importance for the phenomenology associated to the $Z'$ searches. The scalar singlet phenomenology and the dark matter direct detection  constraints involve parameters across the model and will be discussed in detail in later sections.

\section{Anatomy of the electroweak phase transition}
\label{sec:analytical_study}
In this section, we introduce the one loop order scalar potential, based on which, we analyze the thermal history, the phase transition patterns, and the nucleation requirement. With various boundary conditions, derived at zero and finite temperatures, we identify the viable parameter space that can be compatible with the desired thermal history for EWBG. Towards the end of this section, we show numerical results from parameter scannings to support our analytical calculations for the phase transitions.

\subsection{The one loop order scalar potential}
\label{sec:ss}
The one-loop order corrections to the scalar potential in \eq{tree_t0_cpvio} can be calculated via the Coleman-Weinberg (CW) potential in the $\overline{\rm MS}$ scheme using dimensional regularization as \cite{curtin2018thermal}
\begin{eqnarray}
\begin{aligned}
V_{CW}=\frac{1}{64\pi^2}\Biggl\{&\sum_{B}n_B |m_B(h,s,a)|^4\left[\log\left(\frac{|m_B(h,s,a)|^2}{Q^2}\right)-c_B\right]\\
&-\sum_{F}n_F |m_F(h,s,a)|^4\left[\log\left(\frac{|m_F(h,s,a)|^2}{Q^2}\right)-c_F\right]\Biggr\},
\end{aligned}
\label{eq:vcw}
\end{eqnarray}
where $Q$ is the renormalization scale fixed to be the scale of the top quark mass, $m_B$ and $m_F$
are the field-dependent masses of bosons and fermions \cite{Espinosa:1992kf}, $n_B$ and $n_F$ are the degrees of freedom,  $c_B=3/2~(5/6)$ for scalar (vector) bosons, and $c_F=3/2$.

We consider the physical~EW vacuum at zero temperature to be:
\begin{eqnarray}
(h, s,  a)_{T=0} = (v_H, 0,0).
\end{eqnarray} 
To satisfy the extremum condition at the vacuum at one loop order, we require,
\begin{equation}
\begin{aligned}\begin{cases}
&\partial V|_{T=0, (v_H,0,0)}=\partial V_0+\partial V_{CW}+\partial V_{\rm ct}|_{T=0, (v_H,0,0)}=0\\
&\partial^2 V|_{T=0, (v_H,0,0)}=\partial^2 V_0+\partial^2 V_{CW}+\partial^2 V_{\rm ct}|_{T=0, (v_H,0,0)}={\rm Diag}(m_h^2,m_s^2,m_a^2)\end{cases},
\label{eq:condition}
\end{aligned}
\end{equation}
with the Higgs mass $m_h=\sqrt{2 \lambda_H v^2_H} = \unit[125]{GeV}$, and $m_{a,s}$ fixed to be the tree-level relations in \eq{scalarmass}. We introduce counterterms of the following form: 
\begin{equation}
V_{\rm ct}=-\frac{1}{2}\delta_{\mu_H^2}h^2-\frac{1}{2}\delta_{\mu_s^2}s^2-\frac{1}{2}\delta_{\mu_a^2}a^2+\frac{1}{4}\delta_{\lambda_H}h^4
+(\delta_{\rho_S} S+\text{h.c.})
,
\label{eq:counterterm}
\end{equation}
with coefficients fixed by the first and second order derivatives of the CW potential. Note that 
the Hessian matrix evaluated at the~EW vacuum 
contains an off-diagonal term between $s$ and $a$ due to the dark fermion loop. 
We would
leave out
this loop-suppressed
off-diagonal term when fixing the counterterms, and thus the 
small
mixing effect in the singlet sector on singlet phenomenology.

At a finite temperature $T$, the one-loop thermal correction to the effective potential is given by \cite{quiros1998finite}:
\begin{equation}
V_{\text{1-loop}}^T(h,s,a,T)=\frac{T^4}{2\pi^2}\left[\sum_B n_B J_B\left(\frac{|m_B(h,s,a)|^2}{T^2}\right)-\sum_F n_F J_F\left(\frac{|m_F(h,s,a)|^2}{T^2}\right)\right],
\label{eq:thermal_V}
\end{equation}
where $J_B$ and $J_F$ are bosonic and fermionic thermal functions defined as
\begin{equation}
J_B[\alpha]=\int_{0}^{\infty}dx x^2 \log\left[1-e^{-\sqrt{x^2+\alpha}}\right],
~~
J_F[\alpha]=\int_{0}^{\infty}dx x^2 \log\left[1+e^{-\sqrt{x^2+\alpha}}\right],
\label{eq:jf}
\end{equation}
with $\alpha=|m|^2/T^2$. 
Resummation of higher loop daisy diagrams, that ensures the validity of perturbative expansion near the critical temperature of the phase transition, needs to be included in the full one-loop potential. We employ the~Parwani scheme \cite{Parwani:1991gq} for daisy resummation, replacing the tree-level bosonic squared masses
$|m_B(h,s,a)|^2$
with the thermal corrected squared masses 
$|m_B(h,s,a)|^2+\Pi_i(T)$
in \eq{thermal_V} and \eq{vcw}, where $\Pi_i(T)$ is the self-energy calculated from
\cite{quiros1998finite}. In our model, the $\Pi_i(T)$ functions for the fields involved are as follows
\begin{eqnarray}
\begin{aligned}
&\Pi_{h}(T)=\Pi_G(T)=\left(\frac{3}{16} g_2^2+\frac{1}{16} g_1^2+\frac{1}{4} y_t^2+\frac{1}{2}\lambda_H+\frac{1}{12}\lambda_{SH}\right)T^2,\\
&\Pi_s(T) =\Pi_a(T) =\left(\frac{1}{12}\lambda^2+\frac{1}{3}\lambda_S+\frac{1}{6}\lambda_{SH}\right)T^2,\\
&\Pi_{W^{\pm,3},B}^L(T)=\frac{11}{6}T^2\text{diag}(g_2^2,g_2^2,g_2^2,g_1^2),
\end{aligned}
\end{eqnarray}
where $g_2$ and $g_1$ are the SM $SU(2)$ and $U(1)$ gauge couplings, and $y_t$ is the Yukawa coupling of the top quark with the SM Higgs boson.

\subsection{Zero temperature boundary conditions}
\label{sec:zero_temperature}
\begin{figure}
    \centering
    \includegraphics[width=\textwidth]{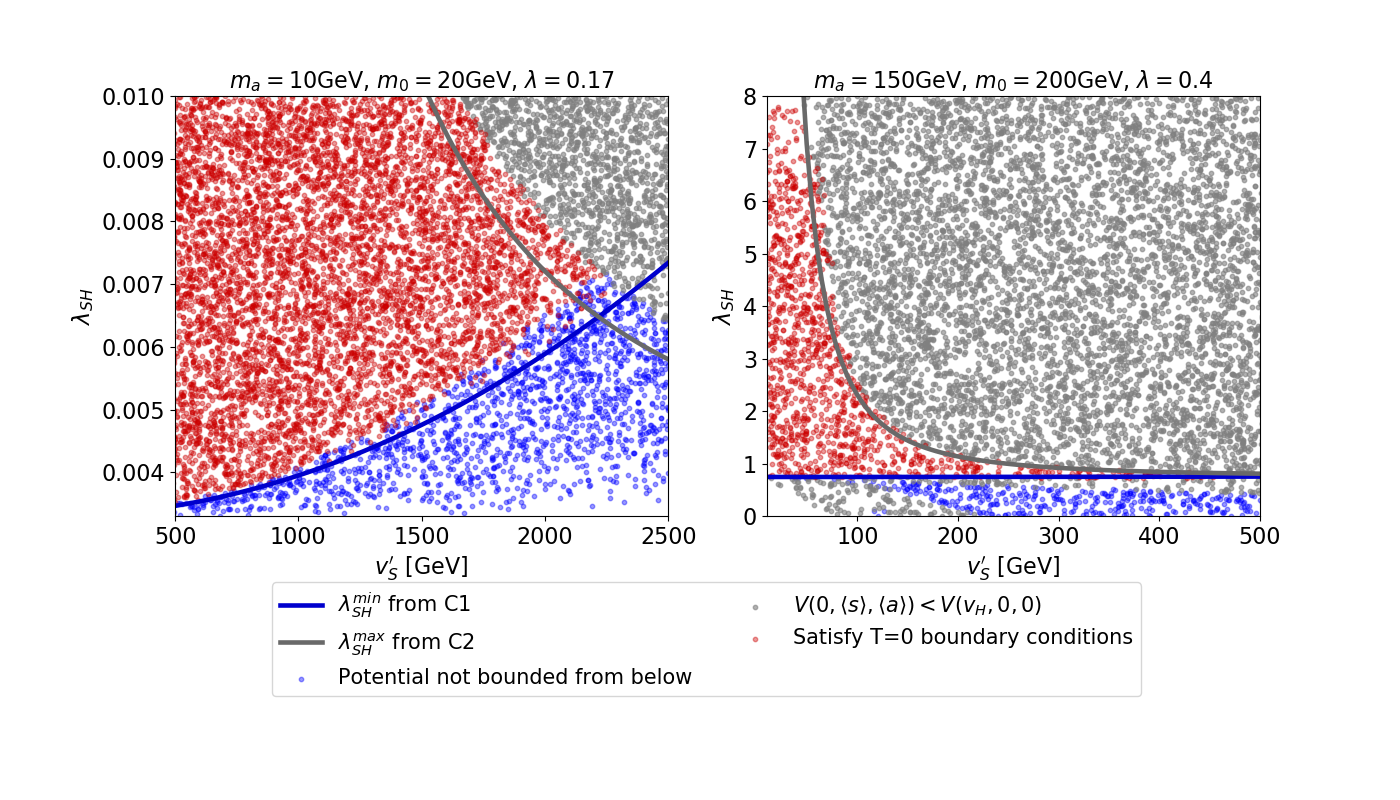}
    \caption{Results of the zero temperature numerical scans at one loop level on the $v_S^\prime-\lambda_{SH}$ plane  
    with fixed parameters (a) $m_a=10$ GeV, $m_0=\unit[20]{GeV}$, $\lambda=0.17$ and (b) $m_a=150$ GeV, $m_0=\unit[200]{GeV}$, $\lambda=0.4$. The benchmarks satisfying zero temperature boundary conditions are shown in red, benchmarks not satisfying BFB are shown in blue and benchmarks where the EW vacuum is not the global minimum are shown in gray.   Analytical conditions \textbf{C1} (loop-level, BFB) and \textbf{C2} (tree-level, global minimum) are shown by the blue and gray lines. The gray points below the blue line, which should also be colored blue, in the right plot have a global minimum different from the EW minimum, despite satisfying \textbf{C2}, due to the significant one loop effects to the small quartic coupling.} 
    \label{fig:T0_vsp_lsh}
\end{figure}
 At zero temperature, the Universe should arrive at the physical vacuum, $(\langle h\rangle, \langle s\rangle, \langle a\rangle)_{T=0} =(v_H, 0, 0)$,
implicating that such a minimum should be the global minimum of the bounded-from-below (BFB) scalar potential. Such requirements imply several boundary conditions on the zero temperature scalar potential, that we summarize here:
 \begin{enumerate}
 \item The stationary point $(v_H,0,0)$ is non-tachyonic;

 On top of requiring a physical Higgs mass $m_h$, this additionally requires non-tachyonic scalar masses $m_s^2 > 0$ and $m_a^2 > 0$. According to \eq{scalarmass}, this at tree level requires
\begin{equation}
\frac{1}{2}\lambda_{SH}v_H^2>\lambda_{S}v_S^{\prime 2}
\label{eq:non-tachyonic}
\end{equation}
 while at one loop level, this is guaranteed by the counterterms;
 
 \item The potential is bounded from below;

At tree level, this is satisfied as we consider the region where all  quartic couplings are positive.
At one loop level, the quartic coupling for the singlet receives negative correction from the dark fermion loop. A necessary condition for the one loop potential to be BFB, can be derived by requiring the tree level coupling being larger than the one loop contribution from the dark Yukawa coupling: 
\begin{equation}
     \lambda_S \gtrsim  \lambda_{S,{\rm BFB}} \equiv \frac{\lambda^4}{16\pi^2}\left[\log\left(\frac{|M_\chi|^2}{Q^2}\right)-\frac{3}{2}\right],
\end{equation}
which can be converted to a lower bound on $\lambda_{SH}$ via \eq{scalarmass}:
\begin{equation}
    \hypertarget{c1}{\textbf{C1}} : \lambda_{SH} \gtrsim \lambda^{\rm min}_{SH}\equiv\frac{2}{v_H^2}\left( \lambda_{S,{\rm BFB}} \cdot v_S^{\prime 2} + m_a^2\right).
    \label{eq:c1}
\end{equation}
In this work we mainly investigate the low energy part of the model, that is related to the~EWPT, around the~EW scale.
Therefore we check numerically that the potential is~BFB up to $10$ TeV before the~UV sector of the complete model factors in.

\item The physical vacuum at $(v_H,0,0)$ is the global minimum of the potential;

 Analysing the stationary points structure of the potential, we start with the tree-level potential given in \eq{tree_t0_cpvio}, focusing on the relations among $m_a$, $v_S^{\prime}$ and $\lambda_{SH}$, neglecting the one loop corrections and hence any constraints on the dark fermion parameters. 
There are four stationary points 
$(h, s, a) = p_i$ for $i=1,\cdots,4$
 (considering only the positive solutions due to $Z_2$ symmetry of the scalar potential at zero temperature), which read
\begin{eqnarray}
\begin{aligned}
&p_1 = (v_H,0,0),~~~~\ p_2 = (0,0,v_S'),~~~~\
p_3 = (0,0,0),\\ &p_4 = \left(\sqrt{\tfrac{2(2\kappa_S^2\lambda_{SH}+\lambda_S \lambda_{SH}v_S^2-2\lambda_S\lambda_H v_H^2)}{\lambda_{SH}^2-4\lambda_S\lambda_H}},0,\sqrt{\tfrac{2\lambda_H (\lambda_{SH}v_H^2-2\lambda_S v_S^2-4\kappa_S^2)}{\lambda_{SH}^2-4\lambda_S \lambda_H}}\right).
\end{aligned}
\label{eq:T0_stationary_points}
\end{eqnarray}
For $p_1$ to be the global minimum, the potential values at $p_2$-$p_4$ must be greater than that at $p_1$. Requiring $V(p_1)<V(p_2)$, we get the following condition:
\begin{eqnarray}
\hypertarget{c2}{\textbf{C2}}:  \lambda_{SH}<\lambda^{\rm max}_{SH} \equiv 2\left(\frac{m_a^2}{v_H^2}+\frac{\lambda_H v_H^2}{v_S^{\prime 2}}\right).
\label{eq:c2}
\end{eqnarray}
$V(p_1)<V(p_3)\Rightarrow -\frac{1}{4}\lambda_H v_H^4<0$ which is automatically satisfied. One can check analytically that, the necessary condition for the existence of $p_4$: $\lambda_{SH}^2-4\lambda_S\lambda_H > 0$, given the fact that $2\lambda_H (\lambda_{SH}v_H^2-2\lambda_S v_S^2-4\kappa_S^2) = 4\lambda_{H}m_a^2 > 0$, guarantees $V(p_1)<V(p_4)$. 

\end{enumerate}

The analytical conditions \hyperlink{c1}{\textbf{C1}} and \hyperlink{c2}{\textbf{C2}} are derived either at tree level or estimated at loop level. In the numerical calculations we impose the above requirements at one loop level. The benchmarks satisfying all conditions will be the ones used for phase transition and EWBG studies. \fig{T0_vsp_lsh} shows the numerical scanning results at one loop level at $T=0$ for two benchmark scenarios  with the parameters: \\
\noindent (a) $m_a=10$ GeV, $m_0=\unit[20]{GeV}$, $\lambda=0.17$, $\theta\in (0,\pi/2)$, $m_s\in (70,200)$ GeV, $v_S^\prime\in (500, 2500)$ GeV, $\lambda_{SH}\in (10^{-5}, 10^{-2})$.\\
\noindent (b) $m_a=150$ GeV, $m_0=\unit[200]{GeV}$, $\lambda=0.4$, $\theta\in (0,\pi/2)$, $m_s\in (155,200)$ GeV, $v_S^\prime\in (10, 500)$ GeV, $\lambda_{SH}\in (10^{-4},8)$.\\
\noindent The analytical conditions  \hyperlink{c1}{\textbf{C1}} and \hyperlink{c2}{\textbf{C2}} are shown as blue and gray lines, respectively. We see that the boundaries set by the analytical conditions agree well with the numerical results, with a few exceptions caused by loop effects.

\subsection{Thermal history}
\label{sec:thermal_history}
As a starting point, to analyse the thermal history, we use the high-temperature expansion of the thermal functions in \eq{jf} to the lowest order of $\alpha$ given by
\begin{equation}
J_B^{\text{high-T}}(\alpha)\approx-\frac{\pi^4}{45}+\frac{\pi^2}{12}\alpha,~~
J_F^{\text{high-T}}(\alpha)\approx \frac{7\pi^4}{360}-\frac{\pi^2}{24}\alpha.
\end{equation}
Such a leading order approximation is sufficient for analytical purposes, as will be shown in the following: a first order phase transition is guaranteed by a tree-level barrier, 
and the thermal barrier arising from the next leading order in the high-temperature expansion
is negligible for achieving a SFOEWPT.

With this expansion, the finite temperature potential can be written as 
\begin{equation}
V(h,s,a,T)=V_0+\frac{1}{2}T^2\left[c_H h^2+c_S (s^2+a^2)+A_s s-A_a a\right]+\text{field-independent terms},
\label{eq:highT_V}
\end{equation}
where coefficients $c_H$, $c_S$, $A_s$ and $A_a$ read
\begin{eqnarray}
\begin{aligned}
c_H&=\frac{3}{16} g_2^2+\frac{1}{16} g_1^2+\frac{1}{4} y_t^2+\frac{1}{2}\lambda_H+\frac{1}{12}\lambda_{SH},\\
c_S&=\frac{1}{12}\lambda^2+\frac{1}{3}\lambda_S+\frac{1}{6}\lambda_{SH},\\
A_s&=\frac{\sqrt{2}}{6}m_0\Re(\lambda_c),\\
A_a&=\frac{\sqrt{2}}{6}m_0\Im(\lambda_c).
\label{eq:thermal_coef}
\end{aligned}
\end{eqnarray}
The first derivatives of 
the effective potential at finite temperature, considering the high temperature expansion in \eq{highT_V}, read
\begin{equation}
    \frac{\partial V}{\partial h}=h\left[-\mu_H^2+\lambda_H h^2+\frac{1}{2}\lambda_{SH}(s^2+a^2)+c_H T^2\right],
    \label{eq:dv_dh}
\end{equation}
\begin{equation}
    \frac{\partial V}{\partial s}=s\left[ -\mu_S^2 + \lambda_S(s^2+a^2)+\frac{1}{2}\lambda_{SH}h^2+2\kappa_S^2+c_S T^2\right] + \frac{1}{2}A_s T^2,
    \label{eq:dv_ds}
\end{equation}
\begin{equation}
    \frac{\partial V}{\partial a}=a\left[ -\mu_S^2 + \lambda_S(s^2+a^2)+\frac{1}{2}\lambda_{SH}h^2-2\kappa_S^2+c_S T^2\right] - \frac{1}{2}A_a T^2.
    \label{eq:dv_da}
\end{equation}
The above equations vanish at the stationary points.
For the analytical understanding of the thermal history that follows, we will fix $\theta = \pi/2$ so that the tadpole coefficients $A_s=0$ and $A_a=\sqrt{2}m_0\lambda/6$. With this simplification, one can prove that all local minima have $s=0$ at arbitrary temperatures, see calculations in \app{singlet_min} \footnote{This choice of $\theta=\pi/2$ would lead to zero CP violation source as will be discussed in Sec. \ref{sec:BA}. Hence this simplification is only for demonstration of the phase transition history, and will not be used for the BAU calculation. In the general numerical scans, we let $\theta$ vary freely in the range $(0,\pi/2)$. However, the thermal history with a general value of $\theta$ follows the same analytic features as 
 $\theta=\pi/2$, that we use as a demonstration.}. Solving for the minima, the thermal history usually goes through several stages as shown by~\fig{thermal_history} and described below:
\begin{figure}[htbp]
\centering
\begin{subfigure}{0.33\textwidth}
\centering
~~\includegraphics[width=\textwidth]{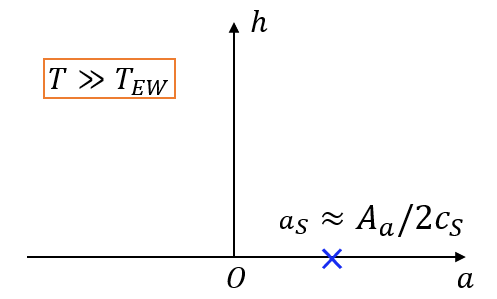}
\caption{}
\label{fig:tha}
\end{subfigure}
\begin{subfigure}{0.33\textwidth}
\centering
~~~\raisebox{0.5mm}{\includegraphics[width=\textwidth]{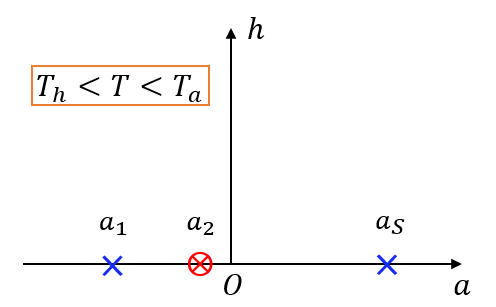}}
\caption{}
\label{fig:thb}
\end{subfigure}\\
\begin{subfigure}{0.33\textwidth}
\centering
\includegraphics[width=\textwidth]{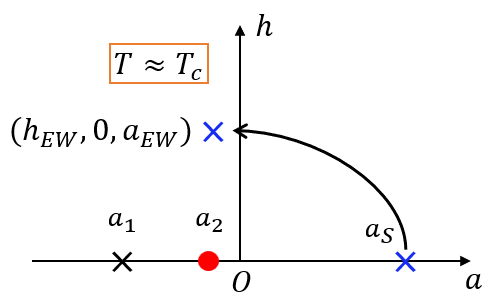}
\caption{}
\label{fig:thc}
\end{subfigure}
\begin{subfigure}{0.33\textwidth}
\centering
~~~\raisebox{1.5mm}{\includegraphics[width=\textwidth]{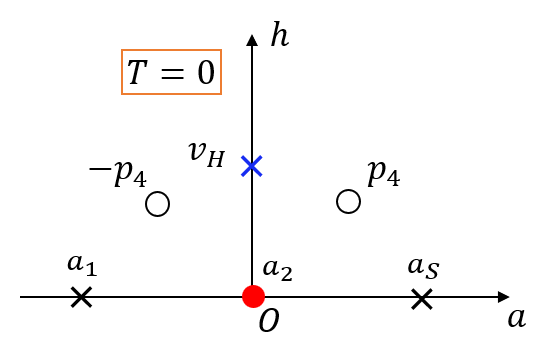}}
\caption{}
\label{fig:thd}
\end{subfigure}
\caption{Patterns of symmetry breaking at different temperatures on the $a-h$ plane with $\langle s\rangle=0$. We use {\color{blue}\Large{$\mathbf{\times}$}} to denote a minimum, {\color{red}$\otimes$} to denote a  saddle point, {\color{red}\Large{$\bullet$}} to denote a maximum, and {\Large{$\mathbf{\times}$}} ($\Circle$) to denote a stationary point that may be minimum (maximum). (a) $T\gg T_{EW}$: The dark Yukawa coupling breaks the $Z_2$ symmetry and induces a non-zero vev for the $a$ field at temperatures much higher than $T_{EW}\approx \unit[140]{GeV}$. (b) Below $T_a$, the temperature  where two additional solutions for the vev of the $a$ field arise, 
there are three branches 
along the direction $h=0$, two of which are minima and one is a saddle point, with the global minimum being $(0,0,a_S)$. When the temperature drops to $T_h$, a new minimum $(h_{EW}$,0,$a_{EW})$ starts to develop from $(0,0,a_2)$. (c) At the critical temperature, the new minimum $(h_{EW}$,0,$a_{EW})$ becomes degenerate with $(0,0,a_S)$. $a_2$ has transformed into a maximum at this temperature. $a_1$ could be a minimum or saddle point depending on the relation between the temperature and the model parameters. (d) At $T=0$, $(v_H,0,0)$ becomes the global minimum and the origin becomes a maximum. There are other potential minima or maxima given by \eq{T0_stationary_points}: $a_2$, $a_1$ and $a_S$ in the plot are the values for the singlet vev at the three stationary points $p_3$ and $\mp p_2$ respectively. Note that we use $-p_i$ to denote the negative solutions.}
\label{fig:thermal_history}
\end{figure}

\begin{enumerate}
    \item At sufficiently high temperatures - well above the EW scale and below the $U(1)_l$ symmetry breaking scale, the thermal corrections dominate, and the symmetry is restored for $h$ and $s$, while $a$ has a non-zero vev that minimizes the effective potential. Hence, 
    the thermal history starts at the vacuum
    \begin{equation} \label{eq:hTa}
     T \gg T_{EW}:\quad  (0,0,a_S)=\left(0,0,\frac{A_a}{2c_S}\right)
    \end{equation}
    as shown in \fig{tha}. We introduce the notation $a_S$ to denote the vev of $a$ at the EW symmetry preserving global minimum. The non-restoration of $Z_2$ symmetry for $a$ is induced by the non-zero $A_a$ from the Yukawa interaction in the dark sector and is an important signature of our model.

\item 
As the Universe cools down but still with the vev along the $h$ direction being zero, the equation for $a$ can be written in the form of a depressed cubic equation
\begin{equation}
    a^3+p(T)a-q(T)=0,
    \label{eq:acubic}
\end{equation}
with
\begin{eqnarray}
\begin{aligned}
p(T)&=\frac{1}{\lambda_S}\left(c_S T^2+m_a^2-\frac{1}{2}\lambda_{SH}v_H^2\right),~~
q(T)&=\frac{A_a T^2}{2\lambda_S}.
\end{aligned}
\end{eqnarray}
Here we have replaced ($-\mu_S^2-2\kappa_S^2$) with ($m_a^2-\tfrac{1}{2}\lambda_{SH}v_H^2$) using \eq{scalarmass}.  The number of real solutions to \eq{acubic} is determined by the sign of its discriminant $D(T)$ with:
\begin{eqnarray}
    D(T) \equiv 4p^3(T)+27q^2(T).
\end{eqnarray}
If $D(T)$ is positive, there will be only one real solution, while, if it is negative, there will be three real solutions to \eq{acubic}. 

 For the model parameter space we investigate, all the quartic couplings are positive, rendering  $c_S > 0$.
Hence, in the case where $\lambda_{SH}<2 m_a^2/v_H^2$, $p(T)$ is always positive, and so is the discriminant $D(T)$, implying  there will only be one real solution to $a$ throughout the thermal history. Because of the $Z_2$ symmetry of the potential in the Higgs direction, i.e.~$h \to -h$,  once an~EW broken stationary point develops, the stationary point $(0,0,a_S)$ necessarily becomes a maximum in the $h$ direction, and a roll-over to the~EW broken stationary point would happen. Note that the inclusion of the thermal cubic term beyond \eq{highT_V} introduces a thermal barrier and the roll-over is promoted to a first-order phase transition, which however is generically weak given the loop-suppression nature of the barrier. We do not investigate this case further.
Thus,
\begin{equation}
    \lambda_{SH}>\frac{2 m_a^2}{v_H^2}.
    \label{eq:lsh_lb}
\end{equation}
is a necessary condition for a SFOEWPT to be induced by a sizable tree-level barrier. This is a weaker constraint than \hyperlink{c1}{\textbf{C1}}, however can be used to down-select the parameter space for the numerical scan.

With \eq{lsh_lb},
$p(T)$ starts out to be positive at high temperatures, 
and turns negative at sufficiently low temperatures, rendering a negative discriminant $D(T)$ and more than one solution to \eq{acubic}.
We define $T_a$ as the temperature at which
\begin{eqnarray}
   D(T_a) = 0.
\end{eqnarray}
$T_a$ can be solved analytically.
To forbid a roll-over to the~EW broken direction until $T_a$, the stationary point $(0, 0, a_S)$ needs to be a minimum along the $h$ direction, leading to 
the necessary condition
\begin{eqnarray} \label{eq:C3}
\hypertarget{c3}{\textbf{C3}}: \frac{\partial^2 V}{\partial h^2} |_{T_a, (0,0,a_S)} = -\mu^2_H +\frac{1}{2} \lambda_{SH} a_S(T_a)^2+ c_H T_a^2> 0,
\end{eqnarray}
with $a_S(T_a) = \sqrt[3]{4q(T_a)}$. This condition provides a lower bound on $\lambda_{SH}$ as a function of $v'_S$. 
Notice that \hyperlink{c3}{\textbf{C3}} also guarantees that the term inside the bracket of \eq{dv_dh} is positive before or at $T_a$ and thus no~EW broken stationary point is developed. 

\item 
Below $T_a$ but above the temperature $T_h$, at which $h$ starts to develop a non-zero vev, two more solutions for $a$ arise, as shown in \fig{thb}. 
The second derivative
    \begin{equation}
        \frac{\partial^2 V}{\partial a^2}=3a^2+p(T)
    \end{equation}
vanish at $a_{\pm} = \pm \sqrt{\frac{|p|}{3}}$ which are the stationary points of the first derivative $\frac{\partial V}{\partial a}$. Thus the three real solutions to \eq{acubic} where $\frac{\partial V}{\partial a} = 0$, denoted as $a_{1, 2, 3}$ with $a_3> a_2> a_1$ should satisfy $a_3 > a_+ > a_2 > a_-> a_1$. 
This also implies that $\partial^2 V/\partial a^2$ is positive for both $a_1$ and $a_3$ and negative for $a_2$. 
To determine the global minimum, we need to compare the potentials at $(0,0,a_1)$ and $(0,0,a_3)$. For the solutions given above, we find that
    \begin{equation}
        V(0,0,a_3)-V(0,0,a_1)=q(T)\lambda_{S} (a_3-a_1)\left[\frac{(a_1+a_3)^2}{4a_1 a_3}-1\right]<0.
    \end{equation}
    Thus right below $T_a$, the Universe would be at the $(0,0,a_3)$, identified as $(0,0,a_S)$, i.e.~the global minimum where the~EW symmetry is preserved.

    \item 
From~\eq{dv_dh}, in order to allow for a non-zero $h$, the following condition should be satisfied as temperature drops:
    \begin{equation}
     -\mu_H^2+\frac{1}{2}\lambda_{SH}(s^2+a^2)+c_H T^2<0.
        \label{eq:mh2neg}
    \end{equation}
    We define $T_h$ as the temperature at which the left-handed side above equals to zero, below which a non-zero $h$ vev starts to develop.
   The stationary point $(0,0,a_2)$ is the first one to satisfy the above condition as the Universe cools down and it smoothly transforms into an EW broken stationary point. As the left-hand side of \eq{mh2neg} also equals to $\partial^2 V/\partial h^2|_{(0,0, a_2)}$, $T_h$ is also the temperature at which $\partial^2 V/\partial h^2|_{(0,0, a_2)}$ starts to be negative. 
   Thus the stationary point $(0,0,a_2)$ transforms from a minimum along the $h$ direction to a maximum as temperature drops below $T_h$. The new stationary point with non-vanishing $h$ will be denoted as $(h_{EW}, 0, a_{EW})$. 
   The second derivative in the $h$ direction is guaranteed to be positive for any EW breaking stationary points given that:
    \begin{equation}
        \frac{\partial^2 V}{\partial h^2}|_{(h_{EW},0,a_{EW})}=2\lambda_H h^2 > 0\ \text{for}\ h\neq 0.
    \end{equation} 
    We expect $(h_{EW}, 0, a_{EW})$ to be the minimum where the Universe settles in after the phase transition. The critical temperature $T_c$ is defined as the temperature at which $V(h_{EW},0,a_{EW}) = V(0,0,a_S)$.
    The~SFOEWPT from $(0,0,a_S)$ to $(h_{EW},0,a_{EW})$ (\fig{thc}) proceeds through bubble nucleation at a slightly lower temperature $T_n$. Since the start and end local minima are separated by a region with $|H|\neq 0$ and $|S|\neq 0$, a barrier is guaranteed by the tree level coupling between $H$ and $S$ proportional to $\lambda_{SH}$. More details on the nucleation condition are to be discussed in \sect{nucl}.

\item Finally, the Universe smoothly deforms into the~EW vacuum $(v_H, 0,0)$ from $(h_{EW}, 0, a_{EW})$ as the temperature drops to zero, with $a_{EW}$ and $a_2$ go to zero, shown in \fig{thd}.
\end{enumerate}
   Analysing the thermal history, in the case of a SFOEWPT, there should be more than one branch of the stationary point solutions in the singlet direction above the phase transition temperature, as shown in \fig{thermal_history}. At the nucleation temperature $T_n$, the false singlet vacuum resides at one branch and tunnels to the true EW vacuum, which develops from another branch.
   Thus, between the false and the true vacuum, a barrier forms in the region where both $|S|$ and $|H|$ are non-zero. Accordingly, we identify the most relevant condition \hyperlink{c3}{\textbf{C3}} for a SFOEWPT, i.e.~\eq{C3}. Numerical study to be shown later agrees very well with such a condition, together with zero temperature boundary conditions, \hyperlink{c1}{\textbf{C1}} and \hyperlink{c2}{\textbf{C2}} we have derived.
Notice that due to the difficulties in solving the cubic equations for the stationary points, we do not present an analytical solution for the critical temperature $T_c$ - the  temperature at which to impose the~SFOEWPT condition. 
However, for light scalar $a$ with mass $m_a<m_h/2$, the coupling $\lambda_{SH}$ is constrained by experiments to be $\lambda_{SH}\lesssim 10^{-2}$. In this scenario, various approximations can be reliably applied on top of the high temperature expansion, which results in an approximate analytic form for $T_c$. With $T_c$ estimated this way, we could provide a lower bound for $\lambda_{SH}$ that agrees better with the numerical results for $m_a<m_h/2$. We describe such an approach and the derived conditions in the following.

\subsection{Critical conditions with light scalars}
\label{sec:lightsingletTc}

For the light singlet scalar with $m_a< m_h/2$, the minimal requirement from Higgs invisible decays imposes that $\lambda_{SH}\lesssim 10^{-2}$. Considering the lower bound for $\lambda_{SH}$ imposed by \eq{lsh_lb}, $m_a$ is bounded to be $m_a\lesssim\unit[17]{GeV}$. The orders of $\lambda_S$ and $v_S^\prime$ consistent with the conditions \hyperlink{c1}{\textbf{C1}}-\hyperlink{c3}{\textbf{C3}} are estimated to be $\lambda_{S}\lesssim 10^{-2}$ and $v_S^\prime\sim \unit[10^3]{GeV}$. Given that we are mainly interested in SFOEWPT, where $h_{EW}(T_c)/T_c> 1$, the temperature involved in the following calculation is bounded to be $T< v_H$. We can solve for $T_c$ analytically with reliable approximations, by evolving the field vevs from their $T=0$ values to higher temperatures. We consider the potential of the form given by \eq{highT_V} to get analytical insights. Based on our analysis above, the SFOEWPT takes place between the two minima $(0,0,a_S)$ and $(h_{EW},0,a_{EW})$, which are smooth deformations of the two stationary points at $T=0$: $p_2 = (0,0,v'_S)$ and $p_1 = (v_H,0,0)$ as in \eq{T0_stationary_points}. The first derivatives of the potential at finite temperatures can be written as Taylor expansions around the stationary points at $T=0$. Expanding around the EW vacuum $(v_H,0,0)$, we have
\begin{eqnarray}
\begin{aligned}
\partial_h V(h,0,a,T)= &m_{h}^2 (h-v_H)+3\lambda_H v_H (h-v_H)^2+\frac{1}{2}\lambda_{SH}v_H a^2+c_H v_H T^2\\
& +\sum_{m+n+p = 3}\frac{1}{m!n!p!}\partial^{1+m}_h\partial^n_a\partial^p_T V|_{(v_H,0,0,T=0)}(h-v_H)^m a^n T^p,
\label{eq:dVdhEW}
\end{aligned}
 \end{eqnarray}
\begin{eqnarray}
 \begin{aligned}
    \partial_a V(h,0,a,T)=&m_{a}^2  a-\frac{1}{2}A_a T^2+\lambda_{SH}v_H a (h-v_H)\\
    &+\sum_{m+n+p = 3}\frac{1}{m!n!p!}\partial^{m}_h\partial^{1+n}_a\partial^p_T V|_{(v_H,0,0,T=0)}(h-v_H)^m a^n T^p,
    \label{eq:dVdaEW}
    \end{aligned}
 \end{eqnarray}
where $m_h^2$ and $m_a^2$ both take their $T=0$ values at the EW vacuum, and the last term in both equations sums over all non-negative integers satisfying $m+n+p=3$. 
Solving \eq{dVdhEW}=0 and keeping only the leading-order terms (see the validity argument in~\app{Tc_estimations}), we get the finite temperature vev of the $h$ field at the EW vacuum
\begin{equation}
    h_{EW}(T)\approx v_H\left(1-\frac{c_H T^2}{2\lambda_H v_H^2}\right).
    \label{eq:hTfinal}
\end{equation}
Substituting this into \eq{dVdaEW}, the leading-order contribution to the finite temperature vev of $a$ at the EW vacuum is found to be
\begin{equation}
    a_{EW}(T)\approx\frac{A_a T^2}{2m_a^2}.
    \label{eq:aEW_T_final}
\end{equation}
 Expanding around the singlet vacuum $(0,0,v'_S)$, 
the first-order derivative along the $a$ direction with $h=s=0$ reads
\begin{eqnarray}
 \begin{aligned}
    \partial_a V(0,0,a,T)=&2\lambda_S (v'_S)^2(a-v'_S)+3\lambda_S v'_S(a-v'_S)^2+\left(c_S v'_S-\frac{1}{2}A_a\right) T^2\\
    &+\sum_{n+p = 3} \frac{1}{n!p!}\partial^{1+n}_a\partial^p_T V|_{(0,0,v_S^\prime,T=0)}(a-v_S^\prime)^n T^p,
    \end{aligned}
 \label{eq:dVdavsp}
\end{eqnarray}
with the last term being a summation over all non-negative integers satisfying $n+p=3$. Solving for the roots and keeping the leading-order terms, we get the finite temperature vev of $a$ at the singlet vacuum (See details in \app{Tc_estimations})
\begin{equation}
    a_S(T)
    \approx v_S^\prime+\frac{T^2}{4\lambda_S v_S^{\prime 2}}\left(A_a-2c_S v'_S\right).
 \label{eq:aS_T_final}
\end{equation}

The critical temperature $T_c$ is when the two stationary points are degenerate 
\begin{equation}
    V(h_{EW},0,a_{EW},T_c)=V(0,0,a_S,T_c),
\end{equation}
which can be written as an expansion around $T=0$ to the second order of the field vevs and temperature:
\begin{eqnarray}
\begin{aligned}
&V(v_{H},0,0,0)+\frac{1}{2}m_h^2|_{(v_H,0,0)}(h_{EW}-v_H)^2+\frac{1}{2}m_a^2|_{(v_H,0,0)}a_{EW}^2+\frac{1}{2}c_H v_H^2 T_c^2\\
&=V(0,0,v'_S,0)+\frac{1}{2}m_a^2|_{(0,0,v'_S)}(a_S-v'_S)^2+\frac{1}{2}(c_S v_S^{\prime 2}-A_a v'_S)T_c^2.
\end{aligned}
\end{eqnarray}
Substituting the field vevs derived above, we get a quadratic equation of $T_c^2$:
\begin{equation}
    \alpha T_c^4+\beta T_c^2 + \gamma=0,
    \end{equation}
with
\begin{eqnarray}
\begin{aligned}
\alpha&=\frac{v_H^2 c_H^2}{2m_h^2|_{(v_H,0,0)}}+\frac{A_a^2}{4\lambda_{SH}v_H^2-8\lambda_S v_S^{\prime 2}}-\frac{\left(A_a - 2c_S v'_S\right)^2}{16\lambda_S v_S^{\prime 2}}\\
\beta&=\frac{1}{2}\left(c_H v_H^2-c_S v_S^{\prime 2} + A_a v'_S\right)\\
\gamma&=\frac{1}{4}\left(-\lambda_H v_H^4+\lambda_S v_S^{\prime 4}\right).
\end{aligned}
\end{eqnarray}
$T_c$ can be solved as 
\begin{equation}
    T_c=\sqrt{\frac{-\beta+\sqrt{\beta^2-4\alpha\gamma}}{2\alpha}}.
    \label{eq:Tcri}
\end{equation}

Similar to condition \hyperlink{c3}{\textbf{C3}}, a necessary condition for SFOEWPT with light singlet scalars at the critical temperature given by \eq{Tcri} reads
\begin{equation}
   \hypertarget{c3p}{\textbf{C3'}}: \frac{\partial^2 V}{\partial h^2}|_{T_c, (0,0,a_S)}>0,
    \label{eq:singlet_min}
\end{equation}
where $a_S$ at $T_c$ can be estimated by 
\eq{aS_T_final}. However, as the temperature dependent term in \eq{aS_T_final} is found to be negligible for light singlets, we will use $a_S(T_c)\approx v'_S$ for the  following calculations. Both \hyperlink{c3}{\textbf{C3}} and \hyperlink{c3p}{\textbf{C3'}} would give a lower bound for $\lambda_{SH}$ as a function of $v'_S$ as shown in \fig{pt_ana}, while \hyperlink{c3p}{\textbf{C3'}} is expected to give a more restrictive bound since $T_c$ is supposed to be lower than $T_a$ for a SFOEWPT.


\subsection{Nucleation}

\label{sec:nucl}
In this section, we derive a  semi-analytical condition for nucleation to proceed.
The potential must be away from the ``thin wall limit'' for the tunneling to happen. The thin wall limit refers to the case where the energy difference between the minima of the potential is significantly smaller than the height of the barrier separating the two phases. The tunneling rate is known to be highly suppressed in such cases. Denoting the singlet phase as $\phi_s$, the EW phase as $\phi_h$ and the barrier as $\phi_b$, this requirement could be formulated as
\begin{eqnarray}
\hypertarget{c4}{\textbf{C4}}: \frac{V(\phi_s)-V(\phi_h)}{V(\phi_b)-V(\phi_h)}>\Delta,
\end{eqnarray}
where $\mathcal{O}(0.1)<\Delta<1$ can be chosen empirically to be consistent with the numerical results \cite{Kozaczuk:2019pet}. This condition should be checked at the temperature where the phase transition completes. Since this is only an approximate condition, we would check this condition at $T=0$ to gain analytical insights. In this case, $\phi_s$, $\phi_h$ and $\phi_b$ would be zero temperature stationary points $p_2$, $p_1$ and $p_4$ given in \sect{zero_temperature}. The ratio $\Delta$ is chosen to be 0.7 to be consistent with the numerical results.

To summarize, we have derived four necessary conditions  (five for $m_a<m_h/2$) analytically from requiring $T=0$ boundary conditions, first order phase transtion and nucleation.
These requirements, which should be checked by numerical calculations at one loop level,
will be discussed in the following section.

\subsection{Numerical scan}
\label{sec:num}
 \begin{figure}
\centering
\resizebox{\textwidth}{!}{
\begin{subfigure}{0.75\textwidth}
\centering
\includegraphics[width=\textwidth]{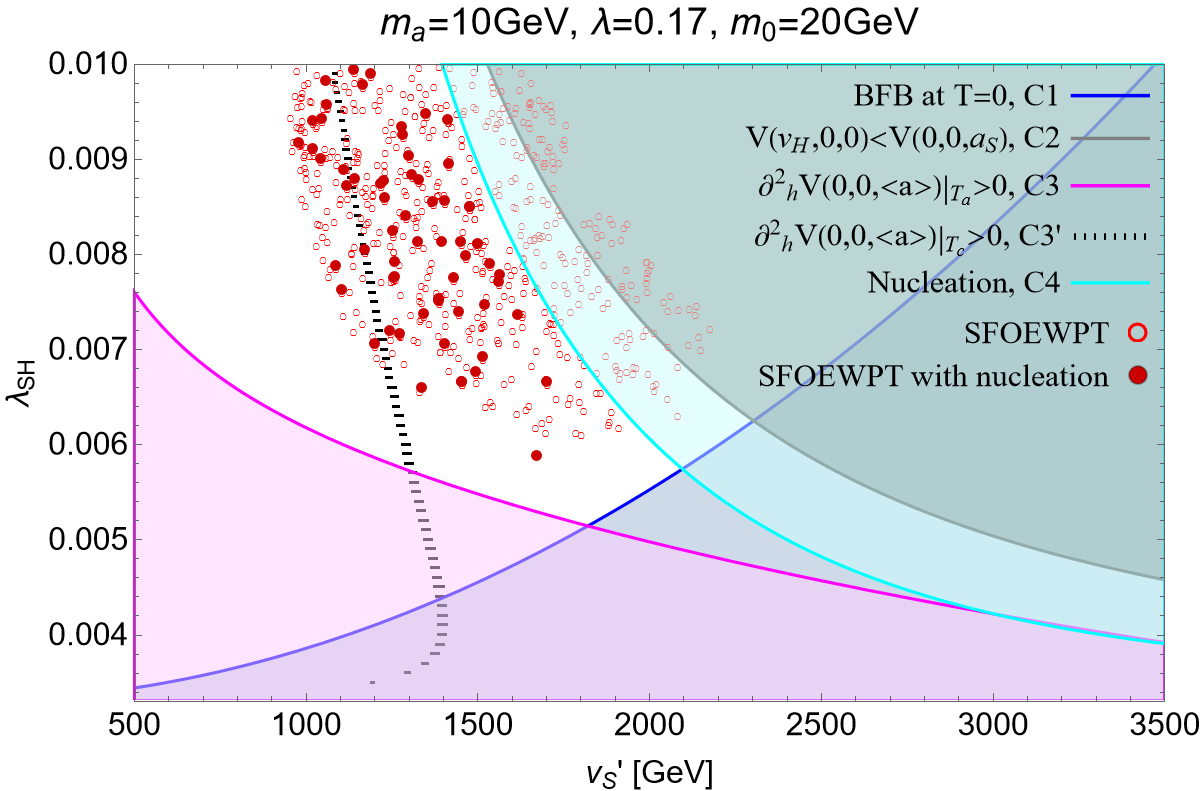}
\caption{}
\end{subfigure}
~~~~~~~\begin{subfigure}{0.7\textwidth}
\centering
\includegraphics[width=\textwidth]{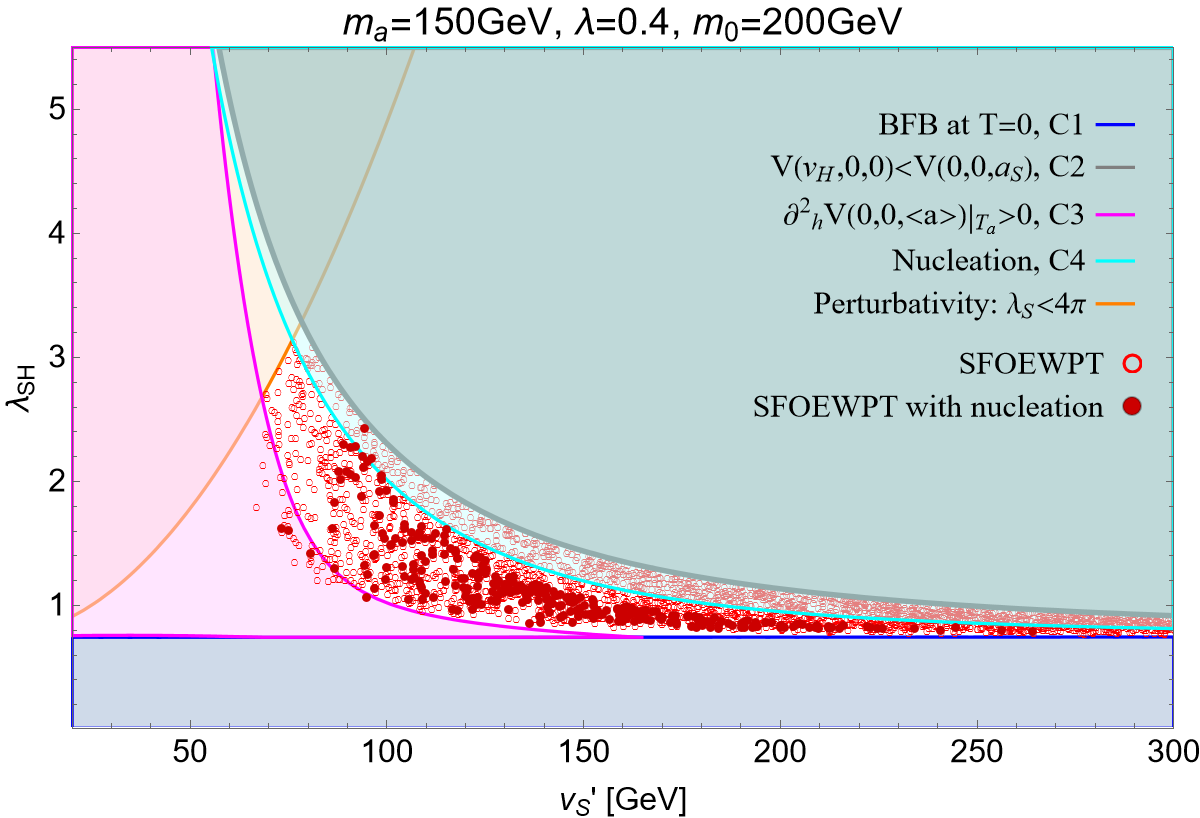}
\caption{}
\end{subfigure}}
\caption{Results of phase transitions on the $v'_S-\lambda_{SH}$ plane for points satisfying the zero temperature boundary conditions for the same two benchmark scenarios  as in \fig{T0_vsp_lsh}:  (a) $m_a=10$ GeV, $m_0=\unit[20]{GeV}$, $\lambda=0.17$ and (b) $m_a=150$ GeV, $m_0=\unit[200]{GeV}$, $\lambda=0.4$. 
 Points satisfying the nucleation and SFOEWPT conditions,  $S_3/T\lesssim 140$ and  $h_{EW}(T_n)/T_n>1$, are shown as red solid circles, and points satisfying the condition $h_{EW}(T_c)/T_c>1$  are shown as red open circles. 
 The analytical bounds derived in \sect{analytical_study},  {\textbf{C1}}-{\textbf{C4}} are shown by the lines.   The shaded regions are excluded by the corresponding conditions.  {\textbf{C3'}}, the condition derived at $T_c$, is shown for $m_a=\unit[10]{GeV}$ only, since the derivation for $T_c$ only applies to the case with $m_a<m_h/2$. The perturbativity requirement $\lambda_S<4\pi$ is shown for $m_a=\unit[150]{GeV}$, while the $\lambda_S$ for $m_a=\unit[10]{GeV}$ is always small enough to satisfy this requirement. The lowest $y$-axis value for $\lambda_{SH}$ for these figures is  set by \eq{lsh_lb}.}
\label{fig:pt_ana}
\end{figure}

We use \texttt{CosmoTransitions} \cite{wainwright2012cosmotransitions} for numerical studies of the phase transitions. This tool starts from the local minima of the scalar potential at zero temperature, tracing the evolution of these minima as temperature increases until it hits some saddle point, which indicates the possible existence of other local minima. At such temperatures, the tool will check the existence of other local minima in the field space close to the saddle point. If it finds one, it will trace back in temperature the evolution of this new minimum until its appearance. With this procedure, the tool can locate all local minima in field space at different temperatures, and thus determine all critical temperatures at which phase transitions may happen.  
However, the phase transitions identified this way by \texttt{CosmoTransitions} are not physically allowed to happen unless they satisfy the following conditions:

\begin{itemize}
\item First order phase transition should proceed via bubble nucleation. By requiring that the expectation value for one bubble to nucleate per Hubble volume is $\sim\mathcal{O}(1)$, one can define the bubble nucleation temperature $T_n$ at which  $S_3(T)/T\sim 140$, with $S_3$ being the 3-dimensional Euclidean action of the instanton integrated over the bubble wall \cite{wainwright2012cosmotransitions}. Around $T_n$, when temperature drops, $S_3(T)/T$ usually drops, and the nucleation will be more efficient.
A first order phase transition can happen only if the nucleation condition given above is satisfied at some point within the phase transition temperature window.

\item For a specific starting phase (the global minimum at higher temperature), if there are more than one eligible ending phases for it to transit to, only the transition occurring at the highest temperature can actually happen, since it occurs prior to all others in the cooling-down history of the Universe. 
\end{itemize}

For a successful EWBG, the first order EWPT needs to be strong enough such that the sphaleron rate inside the bubble is suppressed to preserve the produced baryon asymmetry.
The strength of the phase transition could be measured by the order parameter $h_{EW}(T)/T$ at the phase transition temperature.
We use the following criterion for a SFOEWPT in our numerical scan:
\begin{eqnarray}
\frac{h_{S}(T)}{T}<0.5\ \text{and}\ \frac{h_{EW}(T)}{T}>1, \end{eqnarray}
where $h_{EW}(T)$ and $h_S (T)$ are the Higgs vevs inside and outside the bubble at $T_c$ or $T_n$.
Note that the requirement ${h_{S}(T)}/{T}<0.5$ is to ensure the sphaleron process is efficient outside the bubble. 

We use \texttt{CosmoTransitions} to numerically study the thermal history of the parameter space satisfying zero temperature boundary conditions for the two benchmarks presented in \fig{T0_vsp_lsh}. \fig{pt_ana} shows the parameter space that gives SFOEWPT ({\color{red}\tiny{$\Circle$}}) as defined by $h_{EW}(T_c)/T_c > 1$ as well as the parameter space that satisfies the nucleation condition $S_3/T\lesssim 140$ and SFOEWPT defined by $h_{EW}(T_n)/T_n > 1$ ({\color{red}\small{$\bullet$}}). We also show  the analytical conditions \hyperlink{c1}{\textbf{C1}}-\hyperlink{c4}{\textbf{C4}} (\hyperlink{c3p}{\textbf{C3'}} included for light singlet) in \fig{pt_ana}. For the nucleation condition \hyperlink{c4}{\textbf{C4}}, the empirical parameter $\Delta$ is chosen to be 0.7. Our numerical results agree well with the analytical conditions with only a few exceptions on the boundary. This is due to the difference between the scalar potentials used for the analytical and numerical calculations (\hyperlink{c2}{\textbf{C2}} and \hyperlink{c3}{\textbf{C3}}) as explained in the following, or the approximate nature of the analytical conditions (\hyperlink{c3p}{\textbf{C3'}} and \hyperlink{c4}{\textbf{C4}}). For analytical calculations, we use the high temperature expanded thermal potential up to the leading order plus tree-level zero-temperature potential, while for the numerical calculation, we use the full expression for the thermal potential plus one-loop order zero-temperature potential and daisy resummation in the Parwani scheme.

\section{Baryon asymmetry}
\label{sec:BA}

The baryon asymmetry can be generated during a~SFOEWPT when the Universe 
 tunnels from the electroweak symmetric vacuum to the broken one via bubble nucleation. Both vev of the SM Higgs field and the singlet field $S$ change during the phase transition. 
The real and imaginary parts of the complex singlet across the bubble wall during the phase transition can be modeled by
\begin{equation}
s(z)=\frac{s_0}{2}\left[1+k_s\tanh\left(\frac{z}{L_w}\right)\right],
\end{equation}
\begin{equation}
a(z)=\frac{a_0}{2}\left[1+k_a\tanh\left(\frac{z}{L_w}\right)\right],
\end{equation}
where the coefficients are chosen to match the field vevs. For the $a(s)$ field, with $a_S(s_S)$ and $a_{EW}(s_{EW})$ being the vevs in the EW symmetric and broken phases at the EWPT respectively, we have
\begin{equation}
    a_0=a_S+a_{EW},\quad k_a=\frac{a_S-a_{EW}}{a_S+a_{EW}},
\end{equation}
and analogously for the $s$ field. For general $\theta$ values, $a_S(s_S)$ and $a_{EW}(s_{EW})$ are both non-zero.
$L_w$ is the characteristic scale of the bubble wall width, and we will set it to be a random number in the range $(1/T_n, 10/T_n)$. 
The dark fermion mass can thus be written with explicit spatial coordinate dependence in the rest frame of the bubble wall as
\begin{equation}
M_\chi(z)=m_0+\frac{\lambda}{\sqrt{2}}e^{i\theta}\left[ s(z)+i a(z)\right].
\label{eq:mchi}
\end{equation}
The Yukawa interaction between $\chi_{L,R}$ particle and the $S$ background contributes to the CP violating (CPV) source calculated as \cite{Cline:2006ts}
\begin{eqnarray}
\begin{aligned}
S_{\rm CPV}&=\frac{v_w}{\Gamma_m T_n}\langle\frac{v_z}{2E^2}\rangle\left[|M_{\chi}(z)|^2(\text{arg}M_{\chi}(z))'\right]''\\
&=\frac{v_w}{\Gamma_m T_n}\langle\frac{v_z}{2E^2}\rangle\frac{\lambda}{4 L_w^3}\left[2\sqrt{2}m_0 (s_0 k_s\sin\theta+a_0 k_a\cos\theta)+s_0 a_0 \lambda (k_a-k_s)\right]\\
&\qquad\cdot \left[-2+\cosh\left(\frac{2z}{L_w}\right)\right]\text{sech}^4 \left(\frac{z}{L_w}\right).
\label{eq:scpv}
\end{aligned}
\end{eqnarray}
Non-vanishing $S_{\rm CPV}$ requires non-vanishing $\arg(M_\chi (z))^\prime$. As discussed in \app{singlet_min}, $\theta = \pi/2$ leads to $s_S=s_{EW}=0$ and thus $M_\chi(z)$ is real according to \eq{mchi}, rendering $S_{\rm CPV}=0$.
 With $m_0=0$, both tadpole coefficients $A_a$ and $A_s$ vanish, which also gives $s_S=s_{EW}=0$, rendering $\arg(M_\chi(z))=\theta+\pi/2$ be a constant and thus the $S_{\rm CPV}$ vanishes. 
 Therefore, a non-vanishing $S_{\rm CPV}$ requires $\theta\neq\pi/2$, non-vanishing $m_0$ and $\lambda$. Non-vanishing $S_{\rm CPV}$ also implicitly requires non-vanishing $\kappa_S^2$ as discussed in \sect{mod}. This can also be seen from \eq{dv_ds} and \eq{dv_da}, which show that  $\kappa_S^2=0$ implies $\arg(S)=-\theta$ and thus $M_\chi(z)$ is a real number.

  We comment on the bubble wall velocity we use, which may introduce theoretical uncertainties to the calculation \cite{Cline:2020jre,Dorsch:2021nje,Dorsch:2021ubz,Friedlander:2020tnq,PRD.103.123529}. 
The bubble wall velocity, crucial for the calculation of BAU, is in principle determined by the phase transition dynamics. However, the calculation for the bubble wall velocity is  convoluted due to the non-equilibrium nature of first order phase transitions. In our calculation for BAU, we take a simplified approach that is commonly used in phenomenological studies: treat $v_w$ as a free parameter and truncate the Boltzmann equation to the leading moments \cite{Cline:2006ts}. This approach only applies to subsonic bubble wall velocities, $v_w < 1/\sqrt{3}$, up to theoretical uncertainties. In the following, we shall choose $v_w$ to be a random number in the range $(0,0.5)$.

 Following \cite{carena2019electroweak}, we remind the readers on the generation of the BAU. The CPV source leads to nonzero particle chiral asymmetries in the dark sector defined as:
\begin{eqnarray}
\xi_{\chi_L}(z) = \frac{3}{T^3_n}(n_{\chi_L} - n_{\chi^c_L}),\\
\xi_{\chi_R}(z) = \frac{3}{T^3_n}(n_{\chi_R} - n_{\chi^c_R}),
\end{eqnarray}
with $n_{i}$ being the number density of the corresponding particle. As the sum $\xi_{\chi_{L}}(z) + \xi_{\chi_{R}}(z)$ vanishes due to the conservation of the global $U(1)_\chi$ symmetry, we only need to consider the evolution of $\xi_{\chi_{L}}(z)$ according to the diffusion equation
\begin{equation}
-D\xi''_{\chi_L}-v_w\xi'_{\chi_L}+\Gamma_m(\xi_{\chi_L}-\xi_{\chi_R})=S_{\rm CPV},
\end{equation}
with $D$ being the diffusion constant and $\Gamma_m$ the transport rate.
The solution to this diffusion equation is given by
\begin{equation}
\xi_{\chi_L}(z)=\int_{-\infty}^{\infty}dz_0 G(z-z_0)S_{\rm CPV}(z_0),
\end{equation}
where $G(z)$ is the Green's function.~The chiral asymmetries imply a net $U(1)_l$ charge density near the bubble wall as
\begin{equation}
\rho_l (z)=(q+N_g) (n_{\chi_L} - n_{\chi^c_L})+ q (n_{\chi_R} - n_{\chi^c_R}) = \frac{1}{3}N_g T_n^3\xi_{\chi_L}(z),
\end{equation}
which will yield a Coulomb background of the $Z'$ potential
\begin{equation}
\langle Z'_0 (z)\rangle=\frac{g'}{2M_{Z'}}\int_{-\infty}^{\infty}dz_1 \rho_l (z_1)\exp\left[-M_{Z'}|z-z_1|\right].
\end{equation}
As the singlet vev along the bubble wall is small compared to the $U(1)_l$ symmetry breaking sacle, the change of $M_{Z'}$ along the bubble wall is negligible.

This $Z'$ background effectively acts as a chemical potential $\mu_{L_L}(z)=\mu_{\mathit{l}_R}(z)= g'\langle Z'_0 (z)\rangle$ for the SM leptons and sources the net chiral asymmetry in the SM lepton sector with its thermal equilibrium value given by
\begin{equation}
\Delta n_{L_L}^{EQ}(z)=\frac{2N_g T_n^2}{3}\mu_{L_L}(z)=\frac{2g'N_g T_n^2}{3}\langle Z'_0 (z)\rangle.
\end{equation}
In the presence of sphaleron, which would change lepton and baryon numbers while preserving the SM $B-L$, equal asymmetries would be generated for the SM lepton and baryon numbers, $\Delta n_B=\Delta n_{L_L}$. These asymmetries would evolve towards their equilibrium values following the rate equation given by
\begin{equation}
    \frac{\partial\Delta n_{L_L}(z,t)}{\partial t}=\Gamma_{\rm sph}(z-v_w t)\left[\Delta n_{L_L}^{\rm EQ}(z-v_w t)-\Delta n_{L_L}(z,t)\right].
\end{equation}
The sphaleron rate $\Gamma_{\rm sph}$ at nucleation temperature is considered to be unsuppressed outside the bubble, and exponentially suppressed inside the bubble:
\begin{equation}\Gamma_{\rm sph}(z-v_w t)=
    \begin{cases}
        \Gamma_0, & t<z/v_w\\
        \Gamma_0 e^{-M_{\rm sph}/T_n}, & t>z/v_w
    \end{cases},
\end{equation}
with $\Gamma_0\simeq 10^{-6} T_n$,  $M_{\rm sph}=4\pi h_{EW}(T_n) B/g_2$ being the sphaleron mass inside the bubble, and $B\simeq 1.96$ a fudge factor depending on the Higgs mass and $g_2$ \cite{kuzmin1985anomalous}. The solution to the rate equation is given by
\begin{equation}
    \Delta n_{L_L}=\frac{\Gamma_0}{v_w}\int_0^{\infty} dz \Delta n_{L_L}^{\rm EQ}(z) e^{-\Gamma_0 z/v_w}.
\end{equation}
The observed baryon asymmetry is quantified by the baryon-to-entropy ratio and is measured to be
\begin{equation}
\eta_B=\frac{\Delta n_B}{s} \simeq 0.9 \times 10^{-10}.
\label{eq:observedBAU}
\end{equation}

Having reviewed the dependence of the baryogenesis mechanism on the model parameters, we discuss the large-scale numerical scan we performed  to identify the parameter space that produces the observed baryon asymmetry in \eq{observedBAU}. We  require
 zero temperature boundary conditions, and use \texttt{CosmoTransitions} to identify the parameter space compatible with  nucleation and a SFOEWPT ($v_n/T_n>1$). In addition, we require the $\chi$ dark matter candidate to yield the observed relic density.
The model parameters are highly correlated and restricted by the above requirements and the $\zprime$ search bounds~as summarized below:
\begin{enumerate}

    \item $m_a$: The upper bound for $m_a$ is around $\unit[600]{GeV}$, which is set  by \eq{lsh_lb} in addition with the perturbativity requirement $\lambda_{SH}<4\pi$.  
    \item $m_0$: The dark fermion has to be at least heavier than $2m_s$ to open up the annihilation channel to a pair of singlet scalars. The upper bound is chosen to be 1~TeV to avoid the decoupling of the dark fermion from the thermal history.

    \item $\lambda$ would be chosen to be in the range given by \eq{dmtosinglets} to give the observed dark matter relic density.

    \item $m_s$: Its  difference to $m_a$ is chosen to be in the range $(5,100)$ GeV  to enable the annihilation channel of dark matter to both of the two singlet scalars.

     \item $v'_S$: 
   The upper bound for $v'_S$ comes from the requirement that the lower bound for $\lambda_{SH}$ from \hyperlink{c1}{\textbf{C1}} must be lower than the upper bound for $\lambda_{SH}$ from \hyperlink{c2}{\textbf{C2}},
    which is found to be
    \begin{equation}
        v'_S\lesssim\left(\frac{\lambda_H}{\lambda_{S,{\rm BFB}}}\right)^{\tfrac{1}{4}}v_H.
        \label{eq:vsp_ub}
    \end{equation}
    The upper bound for $v_S^\prime$ is smaller for larger value of $\lambda_{S,{\rm BFB}}$, which occurs for larger dark fermion mass $m_0$ and  stronger dark Yukawa coupling $\lambda$. With heavier singlets, larger $m_0$ and thus larger $\lambda$ are required to obtain the right relic density, hence $v_S^\prime$ is constrained to be smaller. In the numerical scan, the range for $v_S^\prime$ is chosen to be (1, 2.5) TeV for light singlets with $m_{a}<m_h/2$, and (10, 700) GeV for heavy singlets with $m_{a,s}>m_h/2$.

    \item 
    $\lambda_{SH}$: The BFB condition \hyperlink{c1}{\textbf{C1}} sets a lower bound on $\lambda_{SH}$, while the global minimum condition \hyperlink{c2}{\textbf{C2}} sets an upper bound on it. Note that the upper bound is a tree level bound and would be less likely to be fulfilled if the dark fermion loop becomes sizable. 
    \item $g^\prime$: This is chosen to give the observed baryon asymmetry, that is parametrically  proportional to $\sim g'^2/M_{Z'}^2$.
   \item $M_{Z'}$: The parameter region for $M_{Z^\prime} <$ 10~GeV is highly constrained by Kaon and B meson decay \cite{Dror:2017nsg, Dror:2017ehi}, which we will not consider here. On the other hand, for light $m_0$, say $m_0 < M_{Z'}/2$, larger $g^\prime$ is required to achieve the observed BAU, which would be excluded entirely by the $Z'$ searches \cite{carena2019electroweak}. Thus we focus on the case with large $m_0$, especially $m_0> M_{Z'}/2$.
\end{enumerate}

\begin{figure}[hbt!]
\centering
\includegraphics[width=0.7\textwidth]{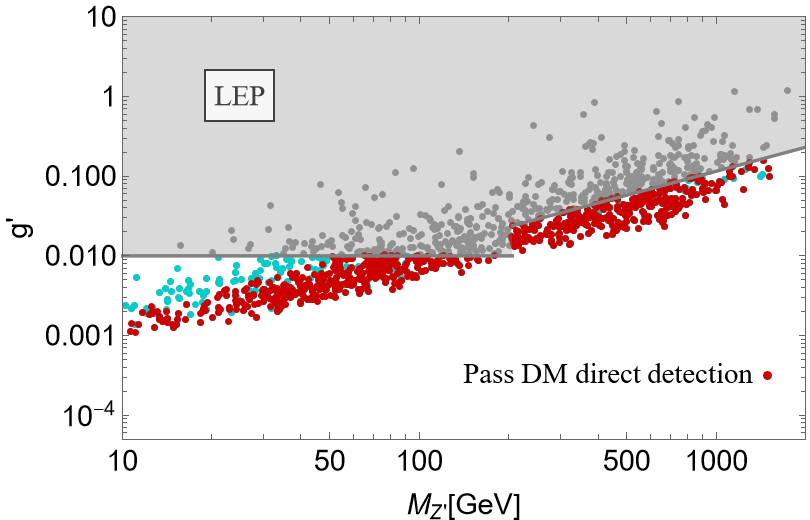}
\caption{Benchmark points producing the observed baryon asymmetry, and the correct dark matter relic density, projected in the dark gauge parameter space defined by  $g'$ and $M_{Z'}$. Gray points are excluded by LEP constraints - the main collider constraint in this mass region. Red and cyan points
survived collider searches, although the latter are excluded by direct dark matter detection searches.}
\label{fig:lepbound}
\end{figure}

The points giving successful nucleation,  observed baryon asymmetry and relic abundance are shown together with the experimental constraints set by LEP \cite{Carena:2004xs,ParticleDataGroup:2018ovx} in~\fig{lepbound}. Gray points in this plot are excluded by LEP constraints, while red and cyan points are not. Cyan points are excluded by the dark matter direct detection bounds as will be discussed in the next section. This plot shows a linear correlation between the logs of the two parameters as expected from the expression for baryon asymmetry, which is proportional to $\sim g'^2/M_{Z'}^2$.

\section{Phenomenology}
\label{sec:phe}

In this section, for the parameter space compatible with the EWBG and the dark matter relic abundance, we 
update the phenomenology on the dark matter direct detection bound on the dark fermion, discuss the search for the singlet scalars at the collider, and study the gravitational wave signatures of the SFOEWPT . We refer the readers to \cite{carena2019electroweak, carena2020dark} for a detailed discussion on the $Z'$ search  and the EDM, while relevant bounds have been applied for the numerical scan.

\subsection{DM direct detection}
\begin{figure}[h]
\centering
\begin{subfigure}{0.45\textwidth}
~~~~~~~~~~~~~~~~~~~~~\includegraphics[width=0.6\textwidth]{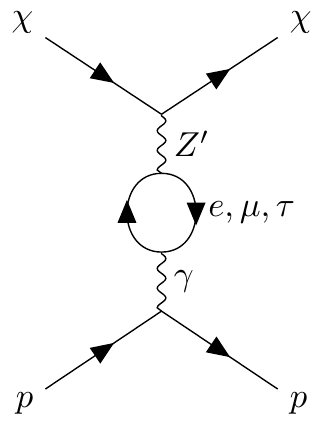}
\end{subfigure}
\begin{subfigure}{0.45\textwidth}
\includegraphics[width=0.6\textwidth]{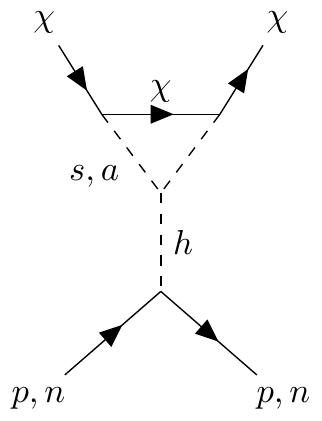}
\end{subfigure}
\caption{Direct detection channels of the dark matter candidate $\chi$.}
\label{fig:dmdirectdetection}
\end{figure}

The most stringent bound from dark matter direct detections for the dark matter mass in our model comes from nuclear recoil experiments. The nuclear recoil
of dark matter occurs through $Z'$ or scalar exchange at one loop order as shown by~\fig{dmdirectdetection} (left). 
The cross section of the scattering process through $Z'$ exchange is given by
\begin{equation}
\sigma_{\chi p\rightarrow\chi p}=\frac{16\alpha^2 \alpha'^2 (q+3/2)^2 \mu_p^2}{81\pi (q^2-M_{Z'}^2)^2}\left[\sum_{l=e,\mu,\tau} f(q^2,m_l)\right]^2,
\label{eq:xsec_dd}
\end{equation}
where $\alpha'=g'^2/(4\pi)$, $\mu_p=m_0 m_p/(m_0+m_p)$. The $f$ function is the loop factor given by
\begin{equation}
f(q^2,m_l)=\frac{1}{q^2}\left[5q^2+12m_l^2-6(q^2+2m_l^2)\sqrt{1-\frac{4m_l^2}{q^2}}\text{arccoth}\sqrt{1-\frac{4m_l^2}{q^2}}+3q^2\log\frac{\Lambda^2}{m_l^2}\right],
\end{equation}
where $q^2=-4\mu^2 v^2$ is the square momentum transfer, with $v\simeq 10^{-3}$ being the typical halo dark matter velocity, $\mu=m_0 m_{\rm Xe}/(m_0+m_{\rm Xe})$, and $\Lambda$ being the renormalization scale chosen to be 1 TeV. In the limit $q^2\rightarrow 0$, which is generically true for our parameter region, $f(q^2, m_l)\sim 3 \log\frac{\Lambda^2}{m_l^2}$.

The scattering process through $h$ exchange shown in~\fig{dmdirectdetection} (right) could also be important as it involves dark matter scattering off both protons and neutrons, and $\lambda, \lambda_{SH}$ can be sizable compared to the $g'$ coupling involved in the $Z'$ exchange channel. The cross section of the scattering process through $h$ exchange in the $q\rightarrow 0$ limit is given by
\begin{equation}
    \sigma^h_{\chi n\rightarrow \chi n} = \frac{y^2_n \lambda^4 \lambda^2_{SH}}{1024\pi^5}\bigg(\frac{\mu_n v_H}{m^2_h m_0}\bigg)^2\bigg|F(\frac{m_s}{m_0})+F(\frac{m_a}{m_0})\bigg|^2,
\end{equation}
with the subscript $n$ in $\mu_n$ representing neutron or proton, and
\begin{equation}
    F(\beta) =-2 + 2(-3+\beta^2)\log\beta -\frac{2\sqrt{-4+\beta^2}(-1+\beta^2)}{\beta}\log\frac{\beta+\sqrt{-4+\beta^2}}{2}.
\end{equation}
The form factor $y_n = 0.3 m_n/v_H$ for the Higgs coupling to nucleon is taken from \cite{Cline:2013gha}. In the numerical calculations, we also take into account the interference term between the two channels in~\fig{dmdirectdetection}. 

\begin{figure}[h]
\centering
\includegraphics[width=0.7\textwidth]{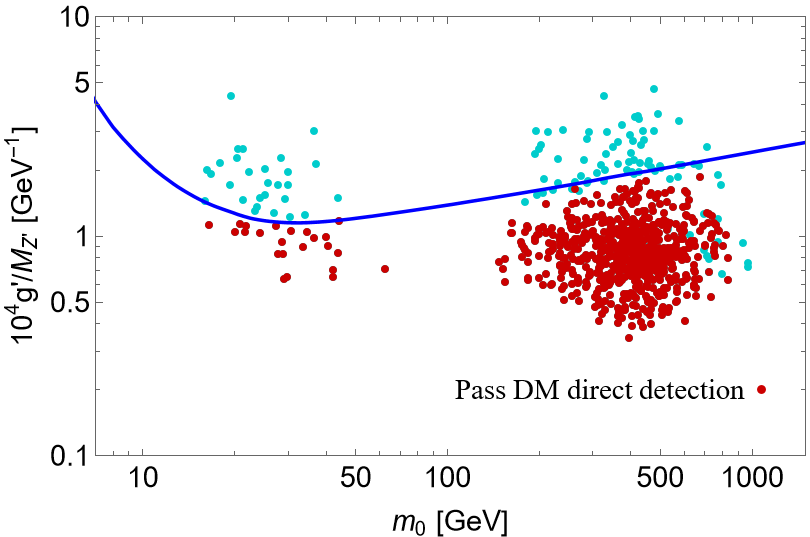}
\caption{Benchmark points producing the observed BAU, relic abundance on the plane of the combined dark gauge parameter ratio $10^4g'/M_{Z'}$ vs the dark fermion mass parameter $m_0$. 
Red points satisfy dark matter direct detection constraints, while cyan points do not. The dark gauge parameters $g'$ and $M_{Z'}$ are calculated to give the observed baryon asymmetry, which is proportional to $\sim g'^2/M_{Z'}^2$. The $Z'$ exchange channel dominates the direct detection cross section in most of the parameter space except for the region where $m_0$ is heavy and $g'/M_{Z'}$ is small. The bound for $g'/M_{Z'}$ (blue line) is derived from the $Z'$ exchange channel cross section, which is proportional to $g'^4/M_{Z'}^4$ at the low energy limit. This bound weakens as the dark matter mass increases, since the direct detection constraint is weaker for heavier dark matter.}
\label{fig:ddbound}
\end{figure}

We show benchmark points (red) on the $m_0$-$g'/M_{Z'}$ plane in \fig{ddbound} that produce the 
observed baryon asymmetry, and satisfy LEP constraints, and pass the current DM direct detection bound from XENON1T \cite{XENON:2018voc}. We also show the upper bound that passes DM direct detection when considering only the $Z'$ channel (blue line) in \fig{ddbound}. According to \eq{xsec_dd}, the direct detection cross section depends on the model parameters $g'$ and $M_{Z'}$ and is proportional to $g'^4/M_{Z'}^4$ in the low $q$ limit. 
As dark matter mass $m_0$ increases from 100 GeV to 1 TeV, its number density decreases by a factor of 10, and the experimental constraints on the direct detection cross section is weakened by one order of magnitude, corresponding to a weaker limit on $g'/M_{Z'}$ by about a factor of 2. For cyan points below the blue line, the contribution to DM direct detection is dominated by the $h$ exchange channel, where $\lambda_{SH}$ can be sizable. This happens at the larger $m_0$ region, where $\chi$ can annihilate to singlet scalars with large $m_{s, a}$ and sizable $\lambda_{SH}$. Future dark matter direct detection from XENONnT will probe the regions with cross section roughly two orders smaller \cite{XENON:2020kmp}. This will probe most of the points in \fig{ddbound} where $g'/M_{Z'}$ is smaller and the dominant contribution to dark matter direct detection comes from the $h$ exchange channel.

\subsection{Singlet scalar searches at LHC}
\begin{figure}[h]
    \centering
    \begin{subfigure}{0.40\textwidth}
    \includegraphics[width=\textwidth]{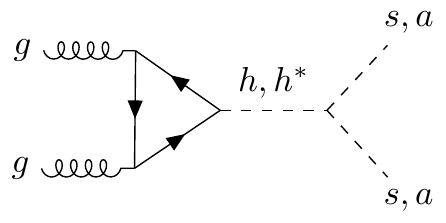}
    \end{subfigure}~~~~~~~~~~
    \begin{subfigure}{0.35\textwidth}
    \includegraphics[width=\textwidth]{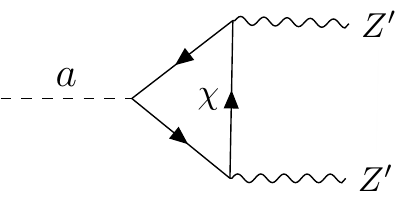}
    \end{subfigure}
    \caption{Main production channel (left) and decay channel (right)  for the light singlet scalar $a$.}
    \label{fig:singletchannel}
\end{figure}
In this section, we explore the collider searches for the new singlet $S$ in our model. 
The physical states from the $S$ field are its real part $s$ and imaginary part $a$, with $a$ lighter than $s$. 
The new scalars $s$ and $a$ can be produced via the Higgs portal. As $S$ has zero vev at zero temperature, both $s$ and $a$ have to be pair-produced via an on-shell or off-shell Higgs boson. In the low energy effective field theory, the coupling between $S$ and $Z'$ can be UV model dependent. In the simple case where $S$ carries $U(1)_l$ charge, the real singlet $s$ can decay to $aZ'$ with $Z'$ further decays to SM leptons. $Z'$ can be on-shell or off-shell depending on the mass difference between $s$ and $a$. We find that $s$ has a decay length mostly smaller than $\unit[0.01]{cm}$ for our parameter space. The signature of $s$ singlets via Higgs production thus contains four leptons and two $a$ singlets in the final states, which we probe with four prompt leptons final state. 

\begin{figure}[h]
    \centering
    \includegraphics[width=\textwidth]{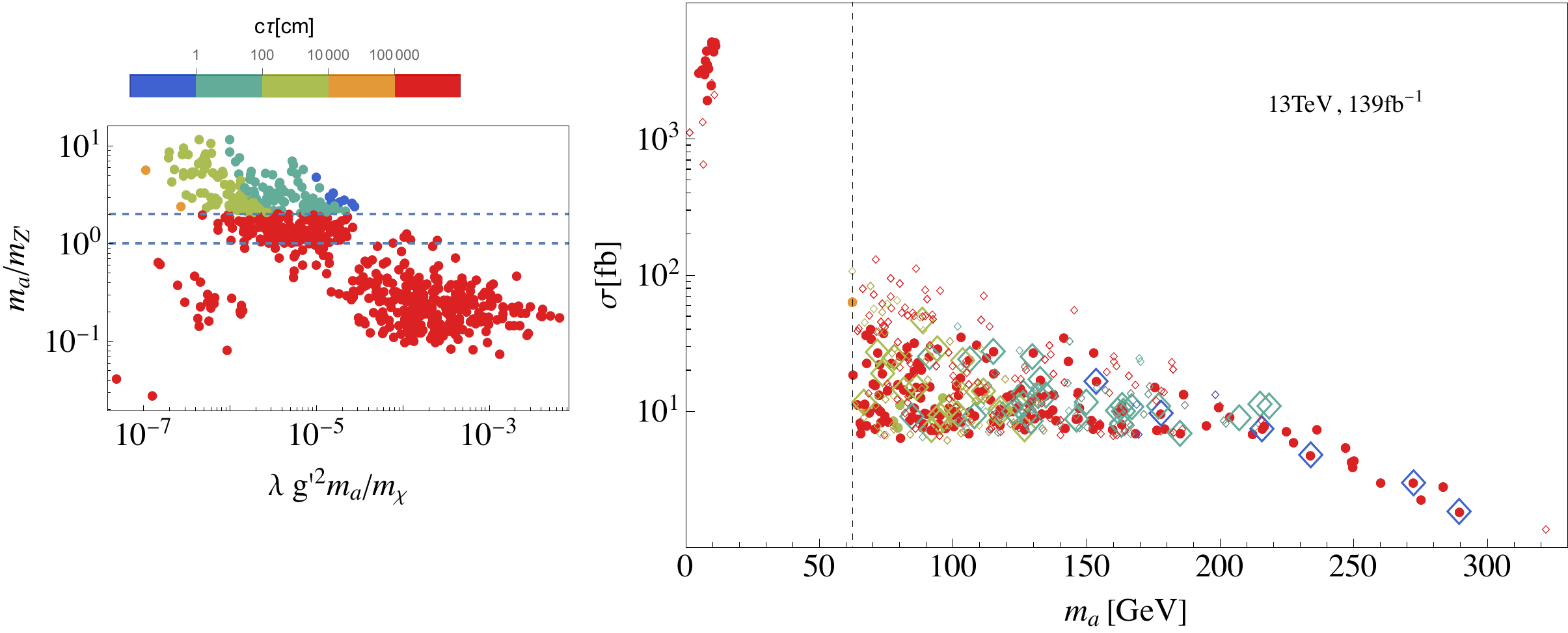}
    \caption{ (left) Lifetime of the real singlet $a$ as a function of $\lambda g'^2 m_a/m_\chi$  ($x$-axis) vs $m_a/m_{Z'}$ ($y$-axis); (right) Constraints from singlet scalar searches at $\unit[13]{TeV}$ LHC with $\mathcal{L} = \unit[139]{fb^{-1}}$. Solid circles pass both $a$ and $s$ search constraints. While most of small open diamonds are excluded by prompt decays of $s$, the larger open diamonds are only excluded by searches for the long-lived particle $a$ singlet. The grey dashed vertical line is the $m_h/2$ threshold. The left and right panels use the same color coding for the lifetime of the singlet $a$.}
    \label{fig:LLP}
\end{figure}

The light singlet $a$ can decay via the dark matter $\chi$ loop to on-shell/off-shell $Z'$ and then further decay to multiple SM leptons, as shown in \fig{singletchannel} (right). As the decay width of this channel is suppressed by both the heavy $\chi$ loop and the small $g'$ coupling, $a$ can be a long-lived particle when produced at colliders. 
The decay length of $a$ is shown in \fig{LLP} (left) indicated by different colors as a function of $\lambda g'^2 m_a/m_\chi$  ($x$-axis) vs $m_a/m_{Z'}$ ($y$-axis). 
The three regions separated by the two horizontal lines in \fig{LLP} (left) contain data points for which the singlet scalar $a$ decays to SM leptons via two on-shell $Z'$s (upper region), one on-shell and one off-shell $Z'$ (middle region) and two off-shell $Z'$s (lower region), respectively.

When $m_{s, a} < m_h/2$, the multi-lepton production cross section at LHC via the singlet scalars is given by $\sigma(gg\to ss(aa)\to l^+l^- + X) \sim \sigma(gg\to h)\times {\rm Br}(h\to ss(aa))$, while when $m_{s, a} \geq m_h/2$, the corresponding cross sectin $\sigma(gg\to ss(aa)\to l^+l^- + X) \sim \sigma(gg\to ss(aa))$ \cite{carena2019electroweak}. We neglect the branching ratio ${\rm {Br}}(Z'\to l^+l^-)\sim \mathcal{O}(1)$ for any leptons. In \fig{LLP}  (right), we present the theoretical estimation of $\sigma(gg\to aa\to l^+l^- + X)$ for all the points that pass all our considerations discussed in the previous sections, as a function of the singlet scalar mass $m_a$.

For the prompt-decay regime ($c\tau < \unit[0.1]{mm}$) with $m_{s, a} < m_h/2$, the bounds on cross section are obtained from the limits at $\unit[13]{TeV}$ on ${\rm Br}(h\to 4l)= 2\times 10^{-5}$ for singlet $s$ with $\unit[139]{fb^{-1}}$ data \cite{Cepeda:2021rql}, while for the singlet $a$, searches for $5l$ final states constrains ${\rm Br}(h\to 5l) \lesssim 10^{-5}$ with $\unit[35.9]{fb^{-1}}$ data \cite{Izaguirre:2018atq}. We further apply these bounds to the parameter space where $m_{s, a} > m_h/2$, \textit{i.e.} singlet $s$ or $a$ is produced via off-shell SM Higgs as an approximation to obtain the points passing collider searches.  

In the case of long-lived singlet $a$ and $s$, we utilize existing searches \cite{CMS:2014hka, CMS:2015pca,Izaguirre:2018atq,CMS:2018uag} for long-lived particles decaying into a pair of electrons or muons, where benchmarks with on-shell Higgs production at different Higgs mass values have been considered.
For the lifetime between $0.1 - \unit[10^4]{mm}$, $a$ and $s$ can be constrained by the searches for lepton pairs with displaced vertices, while for longer lifetimes they can be constrained by searches for muon pairs reconstructed in the muon chamber (see \cite{Alimena:2019zri} and references therein). 
Though bounds on the cross section $\sigma(gg\to ss(aa)\to l^+l^- + X)$ depend on $m_s(m_a)$ and the Higgs boson being on-shell or off-shell, we take the bounds for our parameter regions, as listed in \tab{xsbound}, to be the ones that are satisfied by all benchmarks studied in \cite{CMS:2014hka, CMS:2015pca,Izaguirre:2018atq,CMS:2018uag}. This treatment would result in more stringent bounds for our case due to the fact that our final state leptons can be less energetic than those considered in the benchmark searches. For scalars decaying outside the detector region ($c\tau > \unit[]{km}$), the bounds are obtained from ${\rm Br}(h\to {\rm invisible}) =0.145$ with $\unit[139]{fb^{-1}}$ data \cite{ATLAS:2022yvh}.

We will project these limits to evaluate the reach of the 13 TeV LHC  with luminosity 
$\mathcal{L} = \unit[139]{fb^{-1}}$
 and probe our parameter space. The projection is performed as follows: For the parameter region with $ \unit[0.1]{mm} < c\tau < \unit[]{km}$ where the SM background is negligible, the limit on cross sections roughly scale as $1/\mathcal{L}$. Also with negligible SM backgrounds, the limit on cross section barely depends on the energy of the collisions, hence we can take the results from $\unit[8]{TeV}$ to be approximately valid for $\unit[13]{TeV}$ and only scale the luminosity. For the parameter region with singlet scalars decaying promptly ($c\tau < \unit[0.1]{mm}$) or outside the detector ($c\tau > \unit[]{km}$), the limits on cross section should scale as $1/\sqrt{\mathcal{L}}$ in addition to the energy-dependence from the background cross section. In this case, we consider directly limits from $\unit[13]{TeV}$ and scale them with the luminosity.
The cross section values that can be probed at $95\%$ confidence level at the 13 TeV LHC with luminosity 
$\mathcal{L} = \unit[139]{fb^{-1}}$
 are summarized in \tab{xsbound}, for various values of scalar masses and lifetimes, which are comparable to the bounds recently reviewed in \cite{Cepeda:2021rql} for the scalar singlet with its mass below $m_h/2$.
With more dedicated searches with displaced multi-lepton final states, these bounds can be improved. We leave this study for the future. 

Taking both $a$ and $s$ scalar singlet into consideration, the points that can be probed at 13 TeV LHC with luminosity 
$\mathcal{L} = \unit[139]{fb^{-1}}$ are shown in \fig{LLP} (right) with open diamonds, while points with solid circles require future collider searches. The constraints on different parameter spaces are summarized as follows:
\begin{itemize}
\item $m_{s,a} < m_h/2$, the scalar $s$ can be pair-produced from an on-shell Higgs boson and then promptly decays to $4l+X$. The parameter space will be constrained by the Higgs exotic decay to multi-leptons with its branching ratio constrained to be smaller than $2\times 10^{-5}$ \cite{Cepeda:2021rql}. The small open diamonds to the left of the dashed line in \fig{LLP}(right) fall in this case. Given that: 
\begin{eqnarray}
{\rm{Br}}(h\to ss) = \frac{\lambda^2_{SH}v^2_H}{32\pi m_h \Gamma_h} \sqrt{1-\frac{4m^2_{s}}{m^2_h}},
\end{eqnarray}
with $\Gamma_h = \unit[4.07]{MeV}$ being the SM Higgs total width, $\lambda_{SH}$ is thus constrained to be $\lambda_{SH}\lesssim 10^{-4}$ by Higgs exotic decays. 

\item $m_s>m_h/2$ and $m_a<m_h/2$, the singlet $a$ is produced via on-shell Higgs while $s$ via off-shell Higgs. 
In this scenario, $a$ usually decays outside the detector and the parameter space will be constrained by Higgs invisible decay, which requires $\lambda_{SH}\lesssim 0.01$ ($\sigma < \unit[6380]{fb}$). The solid circle to the left of the dashed line in \fig{LLP}(right) fall in this case.
Searches for singlet $s$ prompt decay give weaker bounds, e.g. $\lambda_{SH}\lesssim 0.07$ for $m_s \sim \unit[70]{GeV}$, with larger upper bound on $\lambda_{SH}$ for larger $m_s$.

\item $m_{s,a}>m_h/2$, both singlet $a$ and $s$ are produced via off-shell Higgs. Searching for  $s$ can be efficient in probing the parameter regions due to the fact that constraints are stronger for particles with shorter lifetimes as observed from \tab{xsbound}. However, as $a$ is lighter than $s$, its production cross section can be much larger, which makes the searches for long-lived $a$ particles more efficient. We present the parameter space that can only be probed by the singlet $a$ search with larger open diamonds. 
\end{itemize}

We can improve on probes of our parameter space through long-lived particle searches by using the detector in~a more creative way, e.g. including the timing information \cite{Liu:2018wte, Liu:2020vur} or considering new LHC auxiliary detectors \cite{Feng:2022inv}. With an expected luminosity increase by roughly a factor of 20, HL-LHC will probe cross sections a factor of 4-20 smaller, allowing us to  probe a broader range of parameter space. For example, consider the searches for singlet $a$ decaying outside the detector, while LHC requires ${\rm Br}(h\to {\rm invisible}) < 0.145$, HL-LHC will constrain ${\rm Br}(h\to {\rm invisible}) < 0.031$, probing all the solid circles with $m_a < m_h/2$ in \fig{LLP}(right).


\begin{table}
\centering
\resizebox{.9\textwidth}{!}{
\begin{tabular}{c|c|c|c}
\hline
    $c\tau[\unit[]{cm}]$ & $\sigma[\unit[]{fb}]$ ($m_{s, a} < m_h/2$) & $\sigma[\unit[]{fb}]$ ($\unit[100]{GeV}>m_{s, a} > m_h/2$)& $\sigma[\unit[]{fb}](m_{s, a} > \unit[100]{GeV}$)\\
    \hline
     $\{0, ~0.01\}$ &0.22, 0.88&0.22, 0.88\footnote{\label{note1} The bounds are approximated as the ones obtained for on-shell SM Higgs ($m_{a,s} < m_h/2$). This is a conservative way to obtain all points passing collider searches.}&0.22, 0.88\footref{note1}\\
     $\{0.01,~0.03\}$&1.44&0.29&0.10\\
     $\{0.03,10\}$&0.72&0.14&0.03\\
     $\{10,100\}$ &1.44&0.29&0.06\\
     $\{100,10^3\}$ &14.4&1.44&0.29\\
     $\{10^3,10^4\}$ &144&14.4&2.88\\
     $\{10^4,10^5\}$ &1440&144&28.8\\
     $\{10^5,-\}$&6380&-&-\\
\hline     
\end{tabular}
}
\caption{Bounds on $(h\to ss(aa)\to l^+l^- + X)$ projected to 13 TeV LHC with luminosity 
$\mathcal{L} = \unit[139]{fb^{-1}}$
 that are satisfied by all the benchmark scenarios (via on-shell Higgs at different values of mass) with similar kinematic regions studied in \cite{CMS:2014hka, CMS:2015pca,Izaguirre:2018atq,CMS:2018uag}. The first number in the first row of each column is the bound obtained by searches for $5$ leptons \cite{Izaguirre:2018atq} which could be applicable to singlet $a$ searches if $a$ decays promptly, while the second number is the bound from Higgs exotic decay \cite{Cepeda:2021rql} to four leptons, applicable to singlet $s$ searches if $s$ decays promptly.
}
\label{tab:xsbound}
\end{table}


\subsection{Gravitational wave signature}

The stochastic gravitational wave generated during an SFOEWPT could become detectable in current and future GW detectors \cite{Caprini:2015zlo}. There are three main sources for GWs from a cosmological first-order phase transition: Collisions of the bubble walls and shocks in the plasma, sound waves in the plasma after the bubble collisions and before the kinetic energy gets dissipated by bubble expansion, and the magnetohydrodynamic (MHD) turbulence in the plasma after bubble collisions. The strengths of these three sources depend on the dynamics of the phase transition, especially the bubble wall velocity $v_w$. We focus on the non-runaway bubble wall scenario with a subsonic terminal velocity $v_w < 1/\sqrt{3}$ for GW calculation, which is compatible with our BAU calculation.
In this scenario, the main contributions to GW signals come from the bulk motion of the fluid, since the energy in the scalar field is negligible. The total power spectrum is a linear combination of contributions from sound waves and MHD turbulence:
\begin{equation}
    h^2\Omega_{GW}\simeq h^2\Omega_{\text{sw}}+h^2\Omega_{\text{turb}},
\end{equation}
with $h$ being the reduced Hubble constant satisfying $H_0 = h\times 100 $km/s/Mpc. Calculations for the GW spectra from the above two sources are presented in \cite{Caprini:2015zlo} as a function of several key parameters. 
Here we comment on such parameters, with $T_{\ast}$ denoting the temperature of the thermal bath when the GW is produced.

\begin{enumerate}
    \item The fraction $\beta/H_{\ast}$, with $\beta$ being the inverse time duration of the phase transition and $H_{\ast}$ being the Hubble constant at temperature $T_{\ast}$, can be evaluated as
    \begin{equation}
   \frac{\beta}{H_{\ast}}\simeq T_{\ast}\frac{d(S_{3}/T)}{dT}|_{T_{\ast}};
    \end{equation}
     $T_{\ast}$ is
     usually taken to be $T_n$ for phase transitions without significant supercooling, as is the case for our parameter space. 

    \item The ratio of the vacuum energy density released during the phase transition to that of the radiation bath, denoted as $\alpha$, is evaluated as
    \begin{equation}
        \alpha=\frac{\rho(\phi_S)-\rho(\phi_{EW})}{\rho_{\text{rad}}^{\ast}},
    \end{equation}
     where $\phi_{EW}$ and $\phi_{S}$ denote the phases inside and outside the bubble wall at the time of the phase transition, and $\rho_{\text{rad}}^{\ast}=g_{\ast}\pi^2 T_{\ast}^4/30$, with $g_{\ast}=106.75$ being the relativistic degrees of freedom in the plasma at $T_{\ast}$;

     \item  The fraction of the total vacuum energy released during the phase transition converted into the bulk motion of the fluid,  is given by
     \begin{equation}
         \kappa_v\simeq\begin{cases}
             \alpha(0.73+0.083\sqrt{\alpha}+\alpha)^{-1}, & v_w\sim 1\\
             v_w^{6/5}6.9\alpha(1.36-0.037\sqrt{\alpha}+\alpha)^{-1}, & v_w\lesssim 0.1
         \end{cases}
     \end{equation}
     For subsonic bubble walls with $v_w<0.5$, the second expression applies. $\kappa_v$ is used to calculate the GW signals from sound waves. The same quantity but for MHD turbulence $\kappa_{\text{turb}}$ is given by
     \begin{equation}
     \kappa_{\text{turb}}=\epsilon\kappa_v,
     \end{equation}
     with $\epsilon$ typically in the range of $5\%-10\%$ \cite{Hindmarsh:2015qta, Caprini:2015zlo}. We use $\epsilon=0.05$ for a conservative evaluation. 
\end{enumerate}

\begin{figure}[h]
    \centering
    \includegraphics[width=0.7\textwidth]{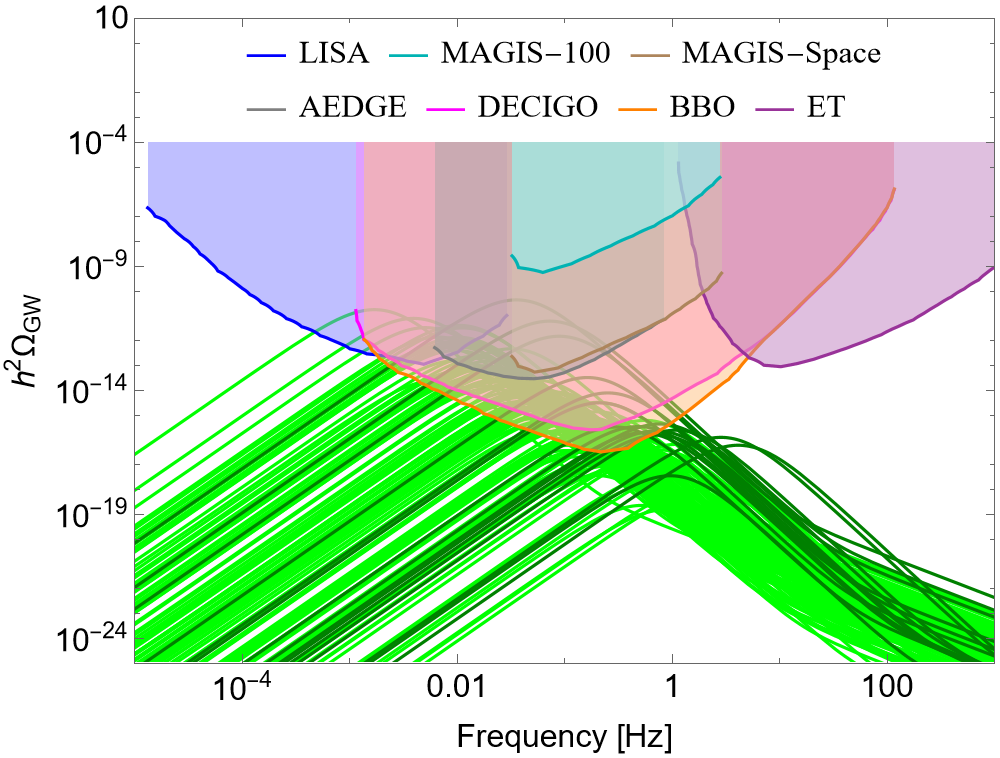}
    \caption{Gravitational wave signals of the SFOEWPT from benchmark points generating the observed baryon asymmetry, the dark matter relic abundance, and satisfying all phenomenological constraints with $m_a<m_h/2$ (dark green curves) and $m_a>m_h/2$ (light green curves). 
The power-law integrated sensitivity curves for LISA, MAGIS-100, MAGIS-Space, AEDGE, DECIGO, BBO, and ET are shown for comparison.}
    \label{fig:gw}
\end{figure}

In \fig{gw}, we show the GW signals calculated for the benchmark points  satisfying all the considerations discussed in previous sections and the comparison to the sensitivities of various proposed GW detectors covering the relevant frequency range \cite{Breitbach:2018ddu}: LISA, DECIGO, BBO, Einstein Telescope (ET), MAGIS-100 and MAGIS-Space \cite{DeLuca:2019llr,MAGIS-100:2021etm}, and AEDGE \cite{AEDGE:2019nxb}. 
The peaks of our GW signals occur between $10^{-3}-10$ Hz, which can be covered by LISA, AEDGE, DECIGO and BBO. GW signals for benchmarks shown in the plot are strong enough to be observed by these detectors based on evaluations presented above. Note that our approach of treating the bubble wall velocity as a free parameter 
would introduce uncertainties to the GW signature calculations, as it should be determined by the specific phase transition dynamics.
There are also alternative calculations \cite{guo2021phase,hindmarsh2021phase,caprini2020detecting} which takes into account the expansion of the Universe and the finite lifetime for the sound waves. Such calculations in general yield weaker signal strengths and lower peak frequencies than our current approach \cite{Caprini:2015zlo}.
More investigation addressing such theoretical uncertainties is needed to be conclusive,
which we will leave for future studies.

\section{Conclusions}
\label{sec:conclusion}
In this work we focus on studying the anatomy of the electroweak phase transition for a recently proposed novel mechanism for electroweak baryogenesis, where the new required source for CP violation resides in a dark sector. Introducing dark CP violation for a successful EWBG evades the stringent constraints imposed by measurements of electron and neutron electric dipole moments. This new EWBG mechanism involves a new fermionic particle $\chi$ and a complex scalar $S$ in the dark sector that together can generate a non-vanishing CPV source. The global lepton number symmetry $U(1)_l$, with $l=e+\mu+\tau$, needs to be promoted to gauge symmetry, with the associated $Z'$ gauge boson that acts as a portal to transfer the CPV source from the dark sector to the SM sector. More in detail, as the Universe undergoes a strong first order phase transition from an EW preserving vacuum to the EW breaking one, the singlet vev is changing along the bubble wall and generates a varying non-vanishing imaginary component for the dark fermion mass, that creates a CPV induced non-zero chiral asymmetry for the dark fermion $\chi$ . This chiral asymmetry is transferred to a chiral asymmetry in the SM lepton sector via the $Z'$ portal. 
The EW sphaleron process in the SM can convert the chiral asymmetry in the SM lepton sector to baryon asymmetry which will be preserved once entering the bubble of EW breaking vacuum.  
The Higgs portal, connecting the SM sector with the dark sector via its interactions with the singlet complex scalar, allows for a  SFOEWPT. Interestingly enough, the fermionic particle $\chi$ introduced can also serve as a dark matter candidate.
In \cite{carena2020dark}, the generated baryon asymmetry at the EWPT was calculated, assuming a sufficiently strong EWPT and that nucleation takes place. Here we perform a meticulous analysis of the phase transition dynamics.  

Exploring the phase transitions, we find distinct thermal histories for the scalar sector: The Yukawa coupling between $\chi$ and $S$, together with the bare mass for $\chi$, breaks the $Z_2$ symmetry ($S\rightarrow -S$) explicitly at tree level and introduces a tadpole term for $S$ at finite temperature. This generically leads to stationary points with non-vanishing vev of $S$ at high temperatures. Later on, the Universe goes through a one-step SFOEWPT to converge to the EW symmetry breaking vacuum. The EWPT proceeding in this way is generically strong due to the barrier between the two vacua induced by the quartic coupling between the complex singlet and the SM Higgs boson. We derive two zero temperature boundary conditions to secure that the potential is bounded from below and the EW breaking vacuum is the global 
minimum. Furthermore, we derive two necessary conditions to achieve SFOEWPT and successful nucleation. The parameter space constrained analytically agrees well with that found by numerical calculations using \texttt{CosmoTransitions}.

Once we identify the  parameter space with successful SFOEWPT, nucleation and dark matter relic abundance for the dark fermion, we evaluate the parameter space that can generate the observed baryon asymmetry while satisfying the current LEP bound from $Z'$ searches. We obtain values of $g'$, the $U(1)_l$ gauge coupling, compatible with the current LEP bounds for $\zprime$ masses between $\unit[10]{GeV}$ and $ \unit[1]{TeV}$. We then examine this scenario via dark matter direct detection searches. Current limit from XENON1T constrains $g'/M_{Z'} \lesssim \unit[10^{-4}]{GeV^{-1}}$ for dark matter masses between 10 GeV to 1TeV. While most of the parameter space is constrained through dark matter scattering with nucleons via the $Z'$ channel, 
there are regions, with sufficiently small $g'$, for which the $h$ channel becomes important.
Smaller scattering cross sections governed by $h$ channel will be probed by future direct detection experiments, e.g., XENONnT.
In this analysis, we assume $\chi$ to be the total dark matter 
relic density, while assuming sub-component dark matter can relax the direct detection bounds opening up more regions of parameter space.

Another interesting probe of this scenario is via searches for the singlet scalars at LHC. The singlet scalars $s$ and $a$ (real and imaginary components of the complex singlet $S$) can be produced via the Higgs portal and subsequently decay to SM leptons through the $\zprime$ portal at tree level ($s$) or via a dark fermion loop ($a$). For the singlet $a$, with the decay width suppressed by both the heavy $\chi$ loop and the small $g'$ coupling, $a$ can be long-lived particles. We calculate the decay lifetime of the singlet scalars for the parameter space leading to successful EWBG and other phenomenological considerations. In the absence of dedicated displaced multi-lepton searches, we utilize previous constraints from displaced di-lepton searches to estimate the regions of parameter space probed by current data at the $\unit[13]{TeV}$ LHC with luminosity $\mathcal{L} = \unit[139]{fb^{-1}}$. We find that the searches for $s$ and $a$ can be complementary to constrain our parameter space: while $s$ can decay relatively faster than $a$, the production cross section for $a$ can be much larger than that for $s$. Thus our parameter space with $s$ not so heavy are mostly constrained by searches for $s$, while with heavy $s$, the parameter space can instead be constrained by searches for $a$ singlet. 
The projection to HL-LHC can be performed straightforwardly, which can probe a broader parameter space for cross sections a factor of 4 to 20 smaller. We expect that powerful probes for long lived scalars  will come from dedicated LHC searches for displaced multi-lepton signals and we encourage the LHC collaborations to look into these topologies. 

We further evaluate the gravitational wave spectra for the parameter space compatible with the observed BAU, dark matter relic abundance and all other phenomenological constraints. The peak of the GW signal for benchmark points arises between $10^{-3}-10$~Hz, and the power spectrum is strong enough to be observed by LISA, AEDGE, DECIGO and BBO. 
We note that there are various theoretical uncertainties associated with the evaluation of the GW signal and the BAU, which need to be further addressed to be conclusive on the simultaneous realization of the EWBG mechanism and strong detectable GW signals.

Summarizing, we have performed a complete analysis, including the anatomy of a sufficiently strong first order EWPT, for a novel mechanism of EWBG with CPV in the dark sector. We show its capability to explain the observed BAU and the dark matter relic abundance.
 The model can be tested by future dark matter direct detection experiments, dedicated search for long-lived particles at the LHC and HL-LHC, as well as future GW laboratory probes. 

\begin{acknowledgments}
We would like to thank Zhen Liu and Yue Zhang for useful discussions and comments.
Fermilab is operated by Fermi Research Alliance, LLC under contract number DE-AC02-07CH11359 with the United States Department of Energy. M.C., Y.-Y.L. and Y.W.~would like to thank the Aspen Center for Physics, which is supported by the National Science Foundation
grant No. PHY-1607611, where part of this work has been done. T.O. is supported by the Visiting Scholars Program of URA. Y.W.~is supported by the U.S.~Department of Energy, Office of Science, Office of High Energy Physics, under Award Number DE-SC0011632.
\end{acknowledgments}

\appendix
\section{Tree level bosonic effective masses}
The zero temperature tree-level potential is given by \eq{tree_t0_cpvio}, and is copied here for completeness:
\begin{eqnarray}
V_0=&&-\frac{1}{2}\mu_H^2 h^2+\frac{1}{4}\lambda_H h^4 +\frac{1}{4}\lambda_{SH}h^2(s^2+a^2) \notag\\
&&-\frac{1}{2}\mu_S^2(s^2+a^2)+\frac{1}{4}\lambda_S(s^2+a^2)^2+\kappa_S^2(s^2-a^2),
\end{eqnarray}
Taking the second derivatives of the potential, we get the masses of the bosons in the model:
\begin{eqnarray}
\begin{aligned}
\partial^2_{G_{i}}V_0&=-\mu_H^2+\lambda_H h^2+\frac{1}{2}\lambda_{SH}(s^2+a^2)\\
\partial_h^2 V_0&=-\mu_H^2+3\lambda_H h^2+\frac{1}{2}\lambda_{SH}(s^2+a^2)\\
\partial_s^2 V_0&=-\mu_S^2+3\lambda_S s^2+\lambda_S a^2+\frac{1}{2}\lambda_{SH}h^2+2\kappa_S^2\\
\partial_a^2 V_0&=-\mu_S^2+3\lambda_S a^2+\lambda_S s^2+\frac{1}{2}\lambda_{SH}h^2-2\kappa_S^2\\
\partial_s \partial_h V_0&=\lambda_{SH} sh,\ \partial_s \partial_a V_0=2\lambda_S sa,\ \partial_a \partial_h V_0=\lambda_{SH}ah,
\end{aligned}
\end{eqnarray}
with $i=1,2,3$.

\section{Local minima with vanishing tadpole coefficient}
\label{app:singlet_min}
In this section, we discuss the solution to the local minima with vanishing tadpole coefficient $A_s$ or $A_a$, which could be obtained by choosing $\theta=\pi/2$, or $\theta=0$ or $m_0=0$ based on \eq{thermal_coef}. 

We first discuss the case with $\theta=\pi/2$, so that $A_s=0$ and $A_a=\sqrt{2}m_0 \lambda/6$. There are two possible sets of solutions to \eq{dv_ds} and \eq{dv_da} with (1) $s=0$, $a\neq 0$, which is the one used for analytical study; (2) $s\neq 0$, $a\neq 0$. For the latter solution, the following identity should be satisfied:
\begin{equation}
    -\mu_S^2+\lambda_{S}(s^2+a^2)+\frac{1}{2}\lambda_{SH}h^2+2\kappa_S^2+c_S T^2=0.
\end{equation}
Using this identity, we can calculate the Hessian matrix in the $s$ and $a$ sector,
\begin{equation}
    \frac{\partial^2 V}{\partial s^2}=2\lambda_{S}s^2,~~~\frac{\partial^2 V}{\partial a^2}=2\lambda_{S}a^2-4\kappa_S^2,~~~\frac{\partial^2 V}{\partial a\partial s}=2\lambda_{S}sa.
\end{equation}
The determinant of this matrix is 
\begin{equation}
    {\rm Det}=-8\lambda_{S}\kappa_S^2 s^2,
\end{equation}
which is negative for positive $\lambda_{S}$ and  $\kappa_S^2$, indicating that this solution would be a saddle point instead of a minimum. Hence all the local minima of the potential with vanishing $A_s$ come with $s=0$. This can be intuitively understood as, $s=0$ is the solution that minimizes the $\kappa_S^2(s^2-a^2)$ term at tree level.

For the case with $\theta=0$ and thus vanishing $A_a$, 
$s\neq 0$ and $a\neq 0$ can be a local stationary point. Following a similar calculation, at such a point,
\begin{equation}
        \frac{\partial^2 V}{\partial s^2}=2\lambda_{S}s^2+4\kappa_S^2,~~~\frac{\partial^2 V}{\partial a^2}=2\lambda_{S}a^2,~~~\frac{\partial^2 V}{\partial a\partial s}=2\lambda_{S}sa.
\end{equation}
The determinant is
\begin{equation}
    {\rm Det}=8\lambda_{S}a^2\kappa_S^2,
\end{equation}
which is positive for positive $\lambda_{S}$ and $\kappa_S^2$. Together with positive diagonal terms, this indicates that a solution with $s\neq 0$ and $a\neq 0$ could be a valid minimum in this case. At finite tempreatures, the Universe would reside at this~CP breaking vacuum if such a stationary point develops to be a global minimum of the potential. One observes that $S_{\rm CPV} \ne 0$ according to \eq{scpv} and successful~EWBG can be realized based on our numerical study. With $\theta =0$, thus all model parameters real, such finite temperature~CP violation is spontaneous and has no zero temperature residue. 

In the case with $m_0=0$, both tadpole coefficients vanish. Following the discussion for vanishing $A_s$, we know the local minima of the potential should have $s=0$.

\section{Critical temperature with light singlet scalars}
\label{app:Tc_estimations}
From the Taylor expansion of the first derivative of the potential around the physical minimum at $T=0$, we can derive the expression for $h_{EW}(T)$ given by \eq{hTfinal}. For the light singlet scalar with mass 
$m_a< m_h/2$, the Higgs and singlet mixing factor is constrained by the singlet searches at colliders, with the minimal requirement $\lambda_{SH}\lesssim 10^{-2}$ from Higgs invisible decays. 
The orders of other parameters in \eq{hTfinal} are estimated to be: $v_H\sim \unit[10^2]{GeV}$, $v_S^\prime\sim \unit[10^3]{GeV}$,
$\lambda_H\sim 10^{-1}$, $\lambda_{S}\sim 10^{-2}$, $c_H\sim 10^{-1}$, $c_S\sim 10^{-2}$ and $A_a\sim \unit[10]{GeV}$, which would be consistent with the conditions \hyperlink{c1}{\textbf{C1}}-\hyperlink{c3}{\textbf{C3}}. Solving \eq{dVdhEW}=0, we get the finite temperature vev of the $h$ field at the EW vacuum:
\begin{eqnarray}
\begin{aligned}
    h_{EW}(T)=&\frac{v_H}{3}\Bigg[2+\sqrt{1-\frac{3\lambda_{SH} a^2}{2\lambda_H v_H^2}-\frac{3c_H T^2}{\lambda_H v_H^2}-\frac{\partial_i^4 V (\Delta x_i)^3}{\lambda_{H}v_H^3}}\Bigg].
    \label{eq:hT}
    \end{aligned}
\end{eqnarray}
where $\partial_i^4 V (\Delta x_i)^3\propto\partial^{1+m}_h\partial^n_a\partial^p_T V|_{(v_H,0,0,T=0)}(h-v_H)^m a^n T^p$. The fourth-order  derivatives $\partial_i^4 V$ are determined by the quartic couplings and the thermal coefficients, among which only the following terms do not vanish
\begin{eqnarray}
    \begin{aligned}
        \partial^4_h V=&6\lambda_{H},~~ \partial^2_a \partial^2_h V=\lambda_{SH}, ~~\partial^2_h \partial^2_T V=2c_H,\\
        &\partial^4_a V=6\lambda_{S}, ~~\partial^2_a \partial^2_T V=2c_S.
    \end{aligned}\label{eq:partial4V}
\end{eqnarray}
We can estimate the orders of $\Delta x_i$ in \eq{hT} based on the known information: (a) SFOEWPT condition, $h_{EW}(T_c)/T_c>1$, indicating that the temperature involved in our calculation satisfies $T< v_H$; (b) $m_h\gg m_a$, indicating that the barrier around the EW vacuum is much higher in the $h$ direction than in the $a$ direction, and thus $\Delta h\ll \Delta a\sim T_c$. Based on these estimations, we see that the leading order term among the three fractional terms under the square root  in \eq{hT} is $\tfrac{3c_H T^2}{\lambda_H v_H^2}$, since the other two terms are either suppressed by the small quartic couplings involving the singlets, or the small change of the $h$ vev. Furthermore, since $T< v_H$, $\tfrac{3c_H T^2}{\lambda_H v_H^2}\ll 1$, \eq{hTfinal} is obtained by keeping only the leading order terms in the Taylor expansion. 

Solving \eq{dVdaEW}=0, we get the finite temperature vev of $a$ at the EW vacuum
\begin{equation}
       a_{EW}(T)=
    \frac{A_a T^2}{2m_a^2}\frac{1-
    \tfrac{\partial_i^4 V (\Delta x_i)^3}{A_a T^2}}{1-\tfrac{\lambda_{SH}c_H T^2}{2\lambda_H m_a^2}},
    \label{eq:aEW_T}
\end{equation}
 with $\partial_i^4 V (\Delta x_i)^3\propto\partial^{m}_h\partial^{1+n}_a\partial^p_T V|_{(v_H,0,0,T=0)}(h-v_H)^m a^n T^p$, and $\partial_i^4 V$ is of order $10^{-2}$ since only the terms involving the derivative with respect to $a$ contribute. The terms $\tfrac{\partial_i^4 V (\Delta x_i)^3}{A_a T^2}$ and $\tfrac{\lambda_{SH}c_H T^2}{2\lambda_H m_a^2}$ are estimated to be of the same order following similar arguments for $h_{EW}(T)$, and thus the second fractional term in \eq{aEW_T} is $\sim\mathcal{O}(1)$ and taken to be 1 for leading order estimation given by \eq{aEW_T_final}.  

Similar estimations can be performed to $a_S(T)$, which is obtained by solving \eq{dVdavsp}=0 as given below
\begin{equation}
    a_S(T)=\frac{v'_S}{3}\left[2+\sqrt{1-\frac{3c_S T^2}{\lambda_S v_S^{\prime 2}}+\frac{3A_a T^2}{2\lambda_S v_S^{\prime 3}}-\frac{\partial_i^4 V (\Delta x_i)^3}{\lambda_S v_S^{\prime 3}}}\right]
    \label{eq:aS_T}
\end{equation}
with $\partial_i^4 V (\Delta x_i)^3\propto\partial^{1+n}_a\partial^p_T V|_{(0,0,v_S^\prime,T=0)}(a-v_S^\prime)^{n} T^p$, and $\partial_i^4 V$ is of order $10^{-2}$. The last fractional term under the square root $\tfrac{\partial_i^4 V (\Delta x_i)^3}{\lambda_S v_S^{\prime 3}}$ is suppressed by the small quartic coupling and $(\Delta x_i/v_S^\prime)^3\ll 1$, and thus is sub-leading compared to the other terms. \eq{aS_T_final} is obtained by keeping only the leading order terms in the expansion. 

\bibliographystyle{JHEP}
\bibliography{EWPT}
\end{document}